\documentclass[fleqn]{aa}

\usepackage{graphicx}
\usepackage{amsmath}
\usepackage[varg]{txfonts}
\usepackage{natbib}
\bibpunct{(}{)}{;}{a}{}{,}

\newcommand{\myemail}{CSandin@aip.de}

\def\rS{SSW08}
\def\rSp{S62}

\def\BD{BD+30$\degr$\,3639}
\def\zgd{\ensuremath{\mathcal{Z}_{\text{GD}}}}
\def\zwr{\ensuremath{\mathcal{Z}_{\text{WR}}}}
\def\kms{\ensuremath{\text{km}\,\text{s}^{-1}}}
\def\teff{\ensuremath{T_{\text{eff}}}}

\def\lsun{\ensuremath{\text{L}_{\odot}}}
\def\msun{\ensuremath{\text{M}_{\odot}}}
\def\te{\ensuremath{T_{\text{e}}}}

\def\na{\ensuremath{n_{\text{A}}}}
\def\ccm{\ensuremath{\text{cm}^{-3}}}

\begin{document}

\title{Hot bubbles of planetary nebulae with hydrogen-deficient winds}
\subtitle{I.\ Heat conduction in a chemically stratified plasma}
\titlerunning{Hot bubbles of PNe with hydrogen-deficient winds}
\author{C.\ Sandin\inst{\ref{inst1}} \and M.\ Steffen\inst{\ref{inst1}} \and D.\ Sch\"onberner\inst{\ref{inst1}} \and U.\ R\"uhling\inst{\ref{inst1},\ref{inst2}}}
\institute{Leibniz-Institut f\"ur Astrophysik Potsdam (AIP), An der Sternwarte 16, D-14482 Potsdam, Germany\label{inst1}
\and
Universit\"at Potsdam, Institut f\"ur Physik und Astronomie, D-14476 Potsdam, Germany\label{inst2}}
\offprints{C.\ Sandin, \email{\myemail}}

\date{Submitted 14 September 2015, Accepted 2 December 2015}

\abstract{Heat conduction has been found a plausible solution to explain discrepancies between expected and measured temperatures in hot bubbles of planetary nebulae (PNe). While the heat conduction process depends on the chemical composition, to date it has been exclusively studied for pure hydrogen plasmas in PNe. A smaller population of PNe show hydrogen-deficient and helium- and carbon-enriched surfaces surrounded by bubbles of the same composition; considerable differences are expected in physical properties of these objects in comparison to the pure hydrogen case. The aim of this study is to explore how a chemistry-dependent formulation of the heat conduction affects physical properties and how it affects the X-ray emission from PN bubbles of hydrogen-deficient stars. We extend the description of heat conduction in our radiation hydrodynamics code to work with any chemical composition. We then compare the bubble-formation process with a representative PN model using both the new and the old descriptions. We also compare differences in the resulting X-ray temperature and luminosity observables of the two descriptions. The improved equations show that the heat conduction in our representative model of a hydrogen-deficient PN is nearly as efficient with the chemistry-dependent description; a lower value on the diffusion coefficient is compensated by a slightly steeper temperature gradient. The bubble becomes somewhat hotter with the improved equations, but differences are otherwise minute. The observable properties of the bubble in terms of the X-ray temperature and luminosity are seemingly unaffected.}

\maketitle

\section{Introduction}\label{sec:introduction}
Space-based observations have shown that the inner cavities of many round or elliptical planetary nebulae (PNe) contain a tenuous and very hot gas that emits predominantly in the soft X-ray domain. From a gas-dynamical point of view, the existence of this gas is expected; the fast central-star wind collides with the slower and denser inner parts of the nebula and becomes shock heated. Given the density and velocity of the stellar wind, the wind shock is adiabatic and the shocked gas is expected to reach temperatures of $10^7$~K or more. The existence of this ``hot bubble'' is essential for the dynamics of a PN. Its pressure is not only sufficient to support the nebula shell against collapse, it also compresses and accelerates the inner parts of the nebula.

All spectral analyses of X-ray emission reveal unexpectedly low characteristic X-ray temperatures of a few $10^6$~K, or even below $10^6$~K. The emission measure is also much higher than expected. \citet{RuChGr.:13} overviews properties of the diffuse X-ray emission from PNe (their Table~3).  The present status and preliminary results of the \textit{Chandra} Planetary Nebula Survey (ChanPlaNS) are presented by \citet{KaMoBa.:12}, \citet{FrMoKa.:14}, and \citet{MoKaBa.:15}.
  
There are two possibilities to remedy the obvious temperature discrepancy between theoretical predictions and observations: (i) heat conduction acts across the interface between the bubble and the nebula as proposed for \ion{H}{ii} regions by \citet{WeMcCa.:77}, and (ii) there is mixing due to hydrodynamical instabilities, which develop at the same interface. Both processes can reduce the electron temperature and increase the density within the bubble, but quantitative results -- i.e., observable quantities that can be tested against real objects -- can only be gained by detailed modeling of the relevant physical processes.

First attempts to model mixing processes by means of two-dimensional radiation-gasdynamics simulations are presented by \citet{StSa:06}, and more recently in more detail by \citet{ToAr:14}. A comparison of the numerical results with observations is still pending. For the first time, \citet[hereafter \rS]{StScWa:08} study the influence of heat conduction on the structure of hot bubbles numerically. We include heat conduction in our one-dimensional radiation-gasdynamics code, which is tailored for studies of the formation and evolution of PNe. In {\rS}, we are able to show that, while the dynamics of a model nebula remains virtually unchanged, the bubble structure and properties such as the X-ray characteristic temperature and luminosity can be well explained by PN models that include heat conduction.
   
The heat conduction depends on the chemical composition of the plasma, as is shown by, e.g., \citet[hereafter \rSp]{Sp:62}. In practical applications, a formalism based on a fully ionized pure hydrogen plasma is mostly used, which is a reasonable approximation for stellar and cosmic plasmas in general.

A small fraction of PNe harbor nuclei with hydrogen-deficient as well as helium- and carbon-enriched surfaces (such as the [WC] Wolf-Rayet objects); these objects also show hydrogen-deficient winds and hot bubbles. A hydrogen-deficient plasma -- that is simultaneously helium- and carbon-rich -- is expected to show heat conduction and line cooling properties that differ considerably from those of a hydrogen-rich plasma. Additionally, because of heat conduction, hydrogen-rich matter from the nebula evaporates into the hydrogen-deficient bubble, where it changes the X-ray and cooling properties of the bubble. A detailed modeling of such hot bubbles that may show a chemical stratification, together with appropriate observations, will be a harbinger for a better understanding of the formation and evolution of hydrogen-deficient central stars of PNe.

A comment is needed on the role of magnetic fields in the context of heat conduction and PNe. The two currently favored mechanisms that could explain the shaping of non-spherical PNe are central-star (CS) binarity and magnetic fields \citep[see, e.g., the review of][]{BaFr:02}. Already very weak magnetic fields inhibit heat conduction efficiently in the direction perpendicular to the field lines \citep[see, e.g., S62;][]{BoBaFr:90,So:94}. Our measurements of the surface magnetic fields in CSPNe \citep{StHuTo.:14} reveal low values of $B < 100$\,G, translating to field strengths of $1$--$100$\,mG in the central hot bubble where heat conduction potentially plays an important role. It turns out that magnetic fields much weaker than $1\,\mu$G would still completely suppress heat conduction perpendicular to the field lines for the physical conditions of hot bubbles of PNe.

In the absence of heat conduction, the diffuse X-ray emission of PNe should be much lower and the characteristic X-ray temperature of the hot bubbles should be higher than observed (e.g., \rS). The observed X-ray properties thus indicate that either any present magnetic fields are non-perpendicular to the radial temperature gradient or some other physical mechanism of the same efficiency as heat conduction is at work \citep[e.g., hydrodynamic instabilities, see][]{ToAr:14}. We restrict our present study to spherical geometry and, as in our previous work, we assume that magnetic fields are absent.

It is the goal of this paper to develop an algorithm for the calculation of heat conduction coefficients for arbitrary chemical compositions and to apply it to the calculation of bubbles with stratified chemical compositions. We overview existing theories in Sect.~\ref{sec:theory}, where we also present a simplified method that suffices for all astrophysical applications. We compare the simpler approach with a more detailed and complex approach in Sect.~\ref{sec:theorycomp} to show that differences are small. Section~\ref{sec:modelcomp} is devoted to a detailed comparison between our new, element-dependent expressions and the pure hydrogen case. We used our one-dimensional radiation-gasdynamics code to model the evolution of a PN with normal composition around a hydrogen-deficient central star with a correspondingly hydrogen-deficient wind. These simulations are guided by properties of a particular object, namely \object{\BD}. Calculated observables are discussed and compared with existing observations in Sect.~\ref{sec:Xray}. We finish the paper with conclusions in Sect.~\ref{sec:conclusions}.

\section{Brief overview of heat conduction theories}\label{sec:theory}
The heat conduction theory we used is based on the Fokker-Planck equation, and was first developed by \citet{CoSpRo:50} and \citet{SpHa:53}, see also {\rSp}. The authors note that the accuracy of the theory is about 5--10\%. Assumptions of this theory are that the gas is fully ionized and that electrons are responsible for all interactions. \citet{De:66,De:67a,De:67b} develops an alternative description of heat conduction (and other transport coefficients) based on a fourth-order series expansion of the Chapman-Enskog-Burnett theory. This alternative description works with partially ionized gases, and is also more computationally demanding than the description of {\rSp}; we compare the two theories in Sect.~\ref{sec:theorycomp}. As in most previous studies of heat conduction, we neglected magnetic fields.

Heat conduction can be described as a diffusion process. The heat flux $\vec{q}$ is a linear function of the gradient of the electron temperature $T_{\text{e}}$,
\begin{eqnarray}
\vec{q}=-D\nabla T_{\text{e}},\label{sandineq}
\end{eqnarray}
where $D$ is the diffusion coefficient. Following {\rS}, we write $D$ as a function of the mean free path $\lambda$ (see below). We used the same approach as \citet{CoMc:77} and multiplied the electron-electron equipartition time $t_{\text{eq}}$ (see Eq.~(5-31) in \rSp) with a characteristic thermal velocity, $v_{\text{char}}$, to calculate a mean free path,
\begin{align}
t_{\text{eq}}&=\frac{3}{4\pi^{1/2}}\frac{m_{\text{e}}^{1/2}k_{\text{B}}^{3/2}T_{\text{e}}^{3/2}}{n_{\text{e}}e^4\ln\Lambda}\quad\text{and}\quad v_{\text{char}}=\left(\frac{3k_{\text{B}}T_{\text{e}}}{m_{\text{e}}}\right)^{1/2},\nonumber\\
\intertext{which is why}
\lambda&=\frac{3\sqrt{3}}{4\pi^{1/2}}\frac{k_{\text{B}}^{2}T_{\text{e}}^{2}}{n_{\text{e}}e^{4}\ln\Lambda}=C_{0}\frac{T_{\text{e}}^{2}}{n_{\text{e}}\ln\Lambda}\,\left[\text{cm}\right]\label{sandinemfp}.
\end{align}
Here, $m_{\text{e}}$ is the electron mass, $k_{\text{B}}$ the Stefan-Boltzmann constant, $n_{\text{e}}$ the electron density, $e$ the electron charge, $\ln\Lambda$ the Coulomb Logarithm, $Z$ the effective (or mean) charge, and $C_{0}=2.62479\times10^{5}\,[\text{cm}^{-2}\text{K}^{-2}]$. Spitzer (1962) expresses $\Lambda$, neglecting shielding by positive ions, as
\begin{eqnarray}
\Lambda\equiv h/p_0=\frac{3}{2ZZ_{\text{e}}e^3}\sqrt{\frac{k_{\text{B}}^3T_{\text{e}}^3}{\pi n_\text{e}}},\label{sandinecl}
\end{eqnarray}
where $h$ is the Debye length, $p_0$ the average closest impact parameter, and $Z_{\text{e}}$ (=1) the electron charge. We used the electron-density weighted effective charge $Z$ of all contributing ions with density $n_i$ and charge $Z_i$,
\begin{eqnarray}
Z=\frac{\sum_in_{\text{e},i}Z_i}{\sum_in_{\text{e},i}}=\frac{\sum_in_iZ_i^2}{n_{\text{e}}}.
\end{eqnarray}
It is necessary to multiply $\Lambda$ by the following factor when $T_{\text{e}}>4.2\times10^5\,\text{K}$ (\rSp),
\begin{eqnarray}
2\alpha c/w=\sqrt{421252/T_{\text{e}}},
\end{eqnarray}
where $\alpha$ is the fine structure constant, $c$ the speed of light, and $w$ the speed of a test electron. We write the logarithm of $\Lambda$ for $T_{\text{e}}\le4.2\times10^5\,$K as
\begin{eqnarray}
\ln\Lambda=9.425+3/2\ln T_{\text{e}}-1/2\ln n_{\text{e}}-\ln Z,\label{eq:smallTe}
\end{eqnarray}
and replace this expression with the following for higher temperatures\footnote{Equation~(3) in {\rS} contains an error; it presents the incorrect constant 22.37. We used the correct value of 15.90 in the actual calculations.}, $T_{\text{e}}>4.2\times10^5\,$K,
\begin{eqnarray}
\ln\Lambda=15.90+\ln T_{\text{e}}-1/2\ln n_{\text{e}}-\ln Z.\label{eq:highTe}
\end{eqnarray}

\begin{table}
\caption{Transport coefficients.}
\label{sandintc}
\tabcolsep=5pt
\begin{tabular}{llllll}
\hline\hline
\noalign{\smallskip}
$Z$ & \multicolumn{1}{c}{1} & \multicolumn{1}{c}{2} & \multicolumn{1}{c}{4} & \multicolumn{1}{c}{16} & \multicolumn{1}{c}{$\infty$}\\\hline\\[-1.8ex]
$\delta_{\text{T}}$ & 0.2252 & 0.3563 & 0.5133 & 0.7977 & 1.0000 \\
$\epsilon$          & 0.4189 & 0.4100 & 0.4007 & 0.3959 & 0.4000 \\
$\zeta_{Z}$            & 1.0000 & 0.7743 & 0.5451 & 0.2092 & 0.0000 \\
\hline
\end{tabular}
\tablefoot{Row~1, the effective charge $Z$; Rows 2 and 3, the transport coefficients $\delta_{\text{T}}$ and $\epsilon$ as defined by \citet{SpHa:53}; Row 4, the factor $\zeta_{Z}$ as defined in Eq.~\ref{sandinefz}.}
\end{table}

\begin{figure}
\centering
\includegraphics{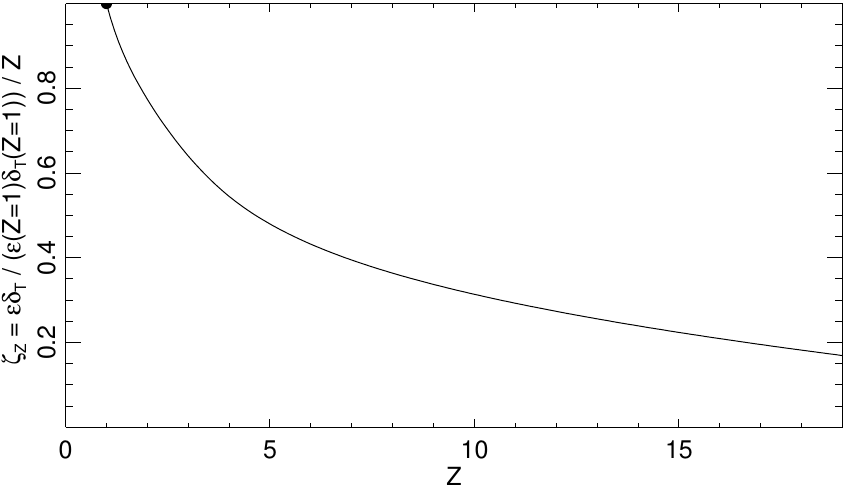}
\caption{The normalized $Z$-dependent term $\zeta_{Z}$ of $\lambda_{Z}$ in $D$ versus the effective charge $Z$.\label{fig1}}
\end{figure}

Writing the expression for the coefficient of heat conduction, we follow {\rSp}, but keep the charge-dependent factors $\epsilon$ and $\delta_{\text{T}}$ in the expression (these are provided in Table~\ref{sandintc}),
\begin{eqnarray}
D&=&20\left(\frac{2}{\pi}\right)^{3/2}\frac{4\pi^{1/2}}{3\sqrt{3}}\frac{k_{\text{B}}n_{\text{e}}}{Z}\left(\frac{k_{\text{B}}T_{\text{e}}}{m_{\text{e}}}\right)^{1/2}\epsilon\delta_{\text{T}}\lambda=\nonumber\\
 &=&D_{1}\lambda_{Z}n_{\text{e}}T_{\text{e}}^{1/2}\,\left[\text{erg}\,\text{s}^{-1}\text{K}^{-1}\text{cm}^{-1}\right],\label{sandineD}\\
\text{where}\\
D_{1}&=&7.45054\times10^{-10}\epsilon_{1}\delta_{\text{T},1}=\,\nonumber\\
&=&7.02856\times10^{-11}\,\left[\text{erg}\,\text{s}^{-1}\text{K}^{-3/2}\text{cm}\right],\nonumber\\
\lambda_{Z}&=&\lambda \zeta_{Z},\,\,\text{and}\,\,\zeta_{Z}=\frac{\epsilon\delta_{\text{T}}}{\epsilon_{1}\delta_{\text{T},1}}\frac{1}{Z}\label{sandinefz}.
\end{eqnarray}
The parameter $\lambda_{Z}$ is the $Z$-dependent mean-free path. For a pure hydrogen composition, $D$ agrees with Eq.~(4) in \rS. We show the charge-dependent term of $D$, $\zeta_{Z}$, versus $Z$ in Fig.~\ref{fig1}; this term excludes the $Z$-dependent component of $\ln\Lambda$, which typically is negligible at high temperatures ($T_{\text{e}}\ga10^6\,$K). For given $T_{\text{e}}$ and $n_{\text{e}}$, $D$ decreases with $Z$, but the heat flux $\vec{q}$ also depends on the temperature gradient, which changes because of the heat conduction. The impact of the effective charge is modest in all cases. For an atmosphere with pure fully ionized carbon ($Z=6$), $D$ is 43 per cent of the value of a pure hydrogen plasma.

The heat flux cannot be larger than the heat content (per unit volume; $E_{\text{th}}$) times a characteristic electron transport velocity \citep[$v_{\text{tr}}$; e.g.,][]{CoMc:77}. Following {\rS}, we define a limiting heat flux $\vec{q}_{\text{sat}}$ using this property,
\begin{eqnarray}
\vec{q}_{\text{sat}}&=&\epsilon\times E_{\text{th}}\times v_{\text{tr}}=\epsilon\times\frac{3}{2}n_{\text{e}}k_{\text{B}}T_{\text{e}}\times\left(\frac{8}{9\pi}\right)^{1/2}\left(\frac{k_{\text{B}}T_{\text{e}}}{m_{\text{e}}}\right)^{1/2}=\nonumber\\
&=&4.28869\times10^{-11}\epsilon n_{\text{e}}T_{\text{e}}^{3/2}.\label{sandines}
\end{eqnarray}
This term agrees with Eq.\ (7) in \citet{CoMc:77} for pure hydrogen when $\epsilon=0.4$. We rewrite Eq.\ (\ref{sandines}) using Eqs.\ (\ref{sandineq}), (\ref{sandineD}), and (\ref{sandinefz}) and get an expression for the mean free path $\lambda_{\text{sat},Z}$ under saturated conditions,
\begin{eqnarray}
\lambda_{\text{sat},Z}=\frac{3\sqrt{3}}{4\pi^{1/2}}\frac{\pi}{40}\frac{\zeta_{Z}Z}{\delta_{\text{T}}}\frac{T_{\text{e}}}{\nabla T_{\text{e}}}\simeq0.057562\frac{\epsilon}{\epsilon_{1}\delta_{\text{T},1}}\frac{T_{\text{e}}}{\nabla T_{\text{e}}}=f_{\epsilon}\frac{T_{\text{e}}}{\nabla T_{\text{e}}},
\end{eqnarray}
where $f_{\epsilon}$ varies very weakly with $Z$ through $\epsilon$, and $f_{\epsilon,1}=0.255604$ for the pure hydrogen case.\footnote{In {\rS}, as in \citet{CoMc:77}, we use $\epsilon_{\infty}$ in Eq.~(\ref{sandines}) and $\epsilon_{1}$ in Eq.~(\ref{sandineD}) to instead get $f_{\epsilon,\infty/1}=0.057562\times\epsilon_{\infty}/\left(\epsilon_{1}\delta_{\text{T},1}\right)=0.244072$.}

The heat conduction terms are included in our \textsc{nebel} code following our approach in {\rS} (see Sect.~2.2.2 therein). Accounting for both regular and saturated heat flux in the same expression, we replaced the mean free path of methods 1 and 2 (see Eqs.~(7) and (8) in {\rS}) with
\begin{align}
\overline{\lambda_{1,Z}}&=\min\left\{f_{\epsilon}\times\Delta r,\lambda_{Z}\right\}\\
\intertext{and}
\frac{1}{\overline{\lambda_{2,Z}}}&=\frac{1}{\lambda_{Z}}+\frac{1}{\lambda_{\text{sat},Z}}=\frac{1}{\lambda_{Z}}+\frac{1}{f_{\epsilon}}\frac{\left|\nabla T_{\text{e}}\right|}{T_{\text{e}}}.\label{sandineh2}
\end{align}
We interpolated the values of $\epsilon$ and $\delta_{\text{T}}$ in Table~\ref{sandintc} using a one-dimensional rational spline \citep{SpMe:90}.

\begin{table}
\caption{Model chemistries used in our theory comparison}
\label{sandintz}
\tabcolsep=5pt
\begin{tabular}{ll}
\hline\hline
\noalign{\smallskip}
Model & Composition (by number)\\[0.0ex]\hline\\[-1.8ex]
$m_1$ & 1/1 H$^+$\\
$m_2$ & 9/10 H$^+$, 1/10 He$^{2+}$\\
$m_3$ & 1/100 H$^+$, 99/100 He$^{2+}$\\
$m_4$ & 1/1 C$^{6+}$\\
$m_5$ & 1/2 C$^{5+}$, 1/2 C$^{6+}$\\
$m_6$ & 1/11 H$^{+}$, 8/11 He$^{2+}$, 2/11 C$^{6+}$\\
$m_7$ & 1/11 H$^{+}$, 8/11 He$^{2+}$, 1/11 C$^{5+}$, 1/11 C$^{6+}$\\
\hline
\end{tabular}
\end{table}

\begin{figure*}
\sidecaption
\includegraphics{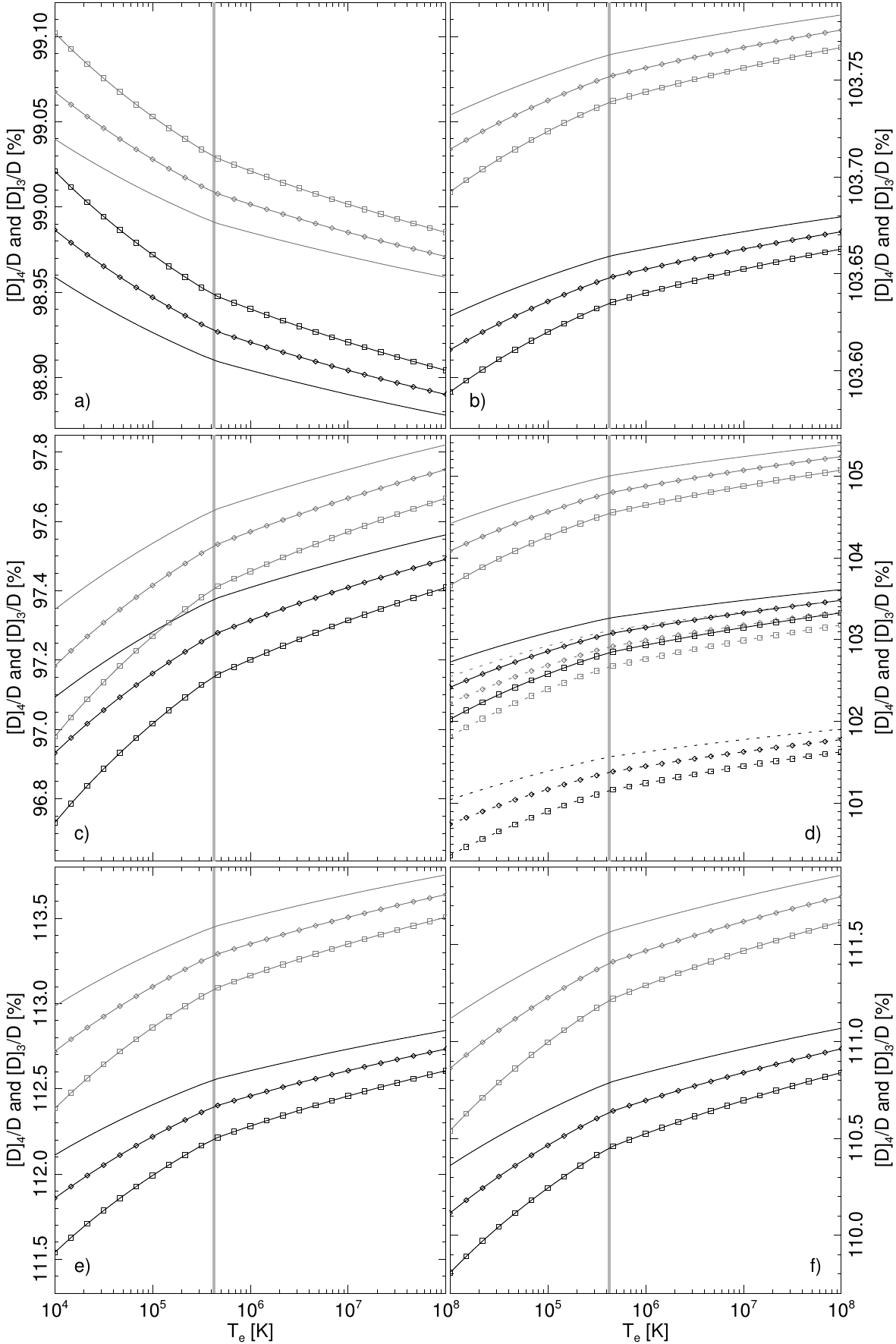}
\caption{Diffusion-coefficient ratios $[D]_4/D$ (gray lines) and $[D]_3/D$ (black lines) drawn versus the electron temperature $T_{\text{e}}$ and the electron density $n_{\text{e}}$. The used densities are $n_{\text{e}}=10^0\,\text{cm}^{-3}$ (no symbol), $n_{\text{e}}=10^2\,\text{cm}^{-3}$ (diamonds), and $n_{\text{e}}=10^4\,\text{cm}^{-3}$ (squares). The six panels show \textbf{a)} $m_1$; \textbf{b)} $m_2$; \textbf{c)} $m_3$; \textbf{d)} $m_4$ (solid) and $m_5$ (dotted); \textbf{e)} $m_6$; \textbf{f)} $m_7$. The vertical gray lines indicate $T_{\text{e}}=4.2\times10^5\,$K.\label{fig2}}
\end{figure*}

\section{Simplified comparison with an alternative theory}\label{sec:theorycomp}
We compare the diffusion coefficient $D$ (Eq.~\ref{sandineD}) with the corresponding fourth- and third-order accurate expressions of \citet{De:67a},
\begin{align}
\left[D\right]_4&=\frac{75n_{\text{e}}^2k_{\text{B}}}{8}\left(\frac{2\pi k_{\text{B}}T_{\text{e}}}{m_{\text{e}}}\right)^{1/2}\left|\begin{matrix}q^{22}&q^{23}\\q^{32}&q^{33}\end{matrix}\right|\times\left|\begin{matrix}q^{11}&q^{12}&q^{13}\\q^{21}&q^{22}&q^{23}\\q^{31}&q^{32}&q^{33}\end{matrix}\right|^{-1}\\
\intertext{and}
\left[D\right]_3&=\frac{75n_{\text{e}}^2k_{\text{B}}}{8}\left(\frac{2\pi k_{\text{B}}T_{\text{e}}}{m_{\text{e}}}\right)^{1/2}q^{22}\times\left|\begin{matrix}q^{11}&q^{12}\\q^{21}&q^{22}\end{matrix}\right|^{-1}.
\end{align}
The lengthy $q^{mp}$ elements ($m,p\in\mathbb{N}$) are provided as functions of Coulomb cross sections in the appendix of \citet{De:67a}; however, there is a typo in those expressions. A numerical check shows that the sign of the second term of q$^{13}$ should be positive as in the other equations (Eq.~a10; `$+8\sum_jn_{\text{e}}n_j\left[\right.\ldots$'). We used the Coulomb cross sections of \citet[see eq.\ 5 therein]{De:67b}. The two theories compared here \citep[{\rSp} and][]{De:67a,De:67b} use different values for the Debye length; the ratio of $\Lambda$ (see Eq.~\ref{sandinecl}) between the two theories is $\Lambda_{\text{Devoto}}/\Lambda_{\text{Spitzer}}=\sqrt{8}/3\approx0.9428$.

To study both hydrogen-rich and hydrogen-deficient plasmas, we used a set of seven models (see Table~\ref{sandintz}). We only considered fully ionized gases and electron-electron interactions; our tests with electron-ion and ion-ion interactions revealed a negligible contribution \citep[this conclusion was reached by comparing the electron-electron terms with the heavy-particle contributions as outlined by][see his Eq.~10]{Ul:70}. The temperature and the density intervals we considered are $10^4\le T\le10^8$\,K and $10^0\le n_{\text{e}}\le10^4\,\text{cm}^{-3}$. While this is a qualitative study, in reality none of the elements heavier than hydrogen is fully ionized at $T_{\text{e}}=10^4\,$K; for studies at such low temperatures we instead refer to \citet{Ul:70}.

The ratios $[D]_4/D$ and $[D]_3/D$ for the different chemistry sets of Table~\ref{sandintz} are illustrated in Fig.~\ref{fig2}. The first impression is that differences are modest, they amount to less than 14\% in the shown examples. Here we compare outcome of mixtures presented in Table~\ref{sandintz}.
\begin{itemize}
\item For a pure hydrogen plasma ($m_1$; Fig.~\ref{fig2}a), the $[D]_4/D$ ($[D]_3/D$) ratio shows small differences 0.900--1.04\% (0.980--1.12\%) lower than $D$. Differences are slightly smaller at higher densities than at lower values.
\item For a hydrogen-helium mixture, differences are larger ($m_2$; Fig.~\ref{fig2}b). The diffusion coefficient $[D]_4$ ($[D]_3$) is 3.68--3.78\% (3.58--3.68\%) higher than $D$.
\item For a helium-dominant mixture ($m_3$; Fig.~\ref{fig2}c), $[D]_4$ ($[D]_3$) is 2.2--3.0\% (2.5--3.3\%) lower than $D$.
\item For a carbon-only mixture ($m_4$ and $m_5$; Fig.~\ref{fig2}d), $[D]_4$ ($[D]_3$) is 4.0--5.3\% (2.2--3.4\%; $m_4$) and 2.0--3.4\% (0.3--1.9\%; $m_5$) higher than $D$.
\item Differences are higher with more WR-like abundances ($m_6$ and $m_7$; Figs.~\ref{fig2}e and \ref{fig2}f). $[D]_4$ ($[D]_3$) is 12.4--13.8\% (11.5--12.9\%; $m_6$) and 10.5--12.0\% (9.81--11.1\%; $m_7$) higher than $D$.
\end{itemize}

We conclude that differences are small in general. The simpler expression in Eq.~(\ref{sandineD}) can be used if there is only hydrogen, or if theory-dependent errors in $D$ of up to 15\% can be tolerated. When the chemistry is more complex and more accuracy is required, the fourth-order expression $[D]_4$ should be used; using this expression instead of $[D]_3$ the diffusion coefficient is up to 2\% higher. Differences in calculation times are negligible with the computing power available today. Considering the many uncertainties in the other parts of this study, we used the simpler (and faster) formulation based on {\rSp} as outlined in Sect.~\ref{sec:theory}.

\section{Comparison of models using the general or the pure-hydrogen heat conduction equations}\label{sec:modelcomp}
Here we study how the weak abundance dependence of the heat conduction is reflected in the outcome of two models that only differ in the use of the heat conduction equation. Model HC uses the chemistry-dependent heat conduction formulation of this paper ($D$, Eq.~\ref{sandineD}), and model HC$_\text{H}$ uses the pure hydrogen formulation of {\rS} ($D_{\text{\rS}}$, their Eq.~4). For given $T_{\text{e}}$ and $n_{\text{e}}$, the diffusion coefficient $D$ of the two formulations differs by the $Z$-dependent factor $\zeta_{Z}$ shown in Fig.~\ref{fig1}.

\subsection{Considerations when setting up models for this study}\label{sec:nebel}
We describe our one-dimensional radiation hydrodynamic (RHD) models of PN envelopes in detail in \citet[and references therein]{PeKiSc.:98,PeScSt.:04}. We emphasize that our models calculate time dependent ionization, recombination, heating, and cooling in a region that extends out to about one parsec away from the CS. The cooling function is composed of the contribution of all considered ions (see Table~\ref{sandint3}), and up to twelve ionization stages are taken into account for every individual ion. Physical input parameters to the models include properties of the coupled CS model, element abundances, and the initial density and velocity structures of matter in the asymptotic giant branch (AGB) envelope.

The purpose of the study is to better model a full system with a Wolf-Rayet star, the preceeding AGB wind, the fast wind, a bubble, and a nebula. To do this we needed to use conditions for these objects that were as realistic as possible. This includes abundances, the AGB evolution, the fast wind evolution, the central-star evolution, and the time when the wind switched from an H-rich state to an H-deficient state. Several of these properties are unknown or poorly known.

\begin{table*}
\caption{Model element abundance distributions. The abundance values are given as both $\epsilon_i=\log(n_i/n_{\text{H}})+12$ and $\beta_i=A_iX_{\text{H}}n_{i}/n_{\text{H}}$.}
\label{sandint3}
\begin{tabular}{rrrrrrrrrr}
\hline\hline\\[-2ex]
\noalign{\smallskip}
\multicolumn{1}{c}{setup}
     & \multicolumn{1}{c}{H}
     & \multicolumn{1}{c}{He}
     & \multicolumn{1}{c}{C}
     & \multicolumn{1}{c}{N}
     & \multicolumn{1}{c}{O}
     & \multicolumn{1}{c}{Ne}
     & \multicolumn{1}{c}{S}
     & \multicolumn{1}{c}{Cl}
     & \multicolumn{1}{c}{Ar}\\[0.0ex]\hline\\[-1.8ex]
\zgd, $\epsilon$ & 12.0 & 11.04 &  8.89 &  8.39 &  8.65 &  8.01 &  7.04 & 5.32 & 6.46 \\
$\beta$ & 68.6 & 29.9$\phantom{0}$ & 0.635 & 0.234 & 0.486 & 0.141 & 239$_{\text{p}}$ & 5.04$_{\text{p}}$ & 78.4$_{\text{p}}$\\
\hline\\[-2ex]
\zwr, $\epsilon$ & 12.0 & 14.50 & 14.10 & 11.70 & 12.99 & 12.32 & 10.35 & 8.63 & 9.77 \\
$\beta$ & 0.0338 & 42.4$\phantom{0}$  & 50.7$\phantom{0}$  &  0.235 &  5.24$\phantom{0}$  &  1.41$\phantom{0}$  & 240$_{\text{p}}$ & 5.07$_{\text{p}}$ & 78.8$_{\text{p}}$\\
\hline
\end{tabular}
\tablefoot{Column 1, model identification symbol, where model parameters are provided in both $\epsilon$ and $\beta$; Cols.~2--10, used abundances of hydrogen, helium, carbon, nitrogen, oxygen, neon, sulfur, chlorine, and argon, respectively. $\beta$ is specified in per cent, except in the last three columns where it is ppm. The hydrogen mass fraction $X_{\text{H}}=\beta_{\text{H}}$. The total $\sum_i\beta_i=1$, which cannot be shown here with only three significant digits.}
\end{table*}

\begin{figure}
\centering
\includegraphics{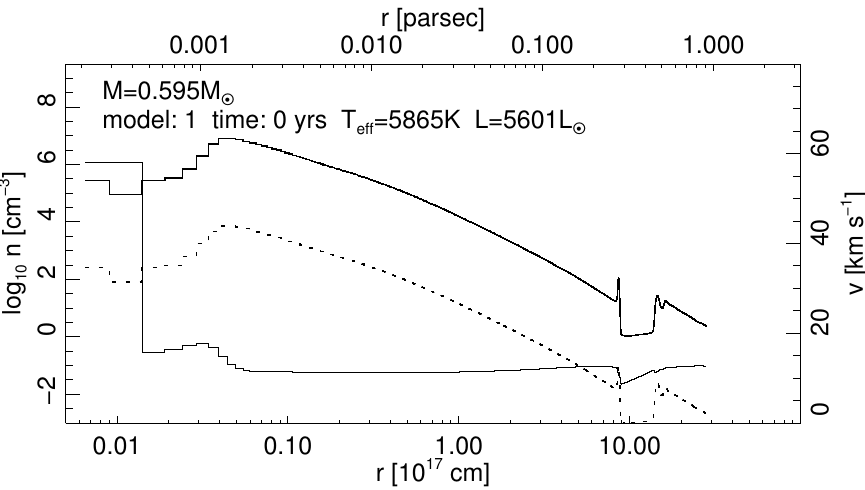}
\caption{Initial AGB wind-envelope structures. The left-side axis is used with heavy particle (electron) densities drawn with a thick solid (dotted) line. The right-side axis is used with flow velocities drawn with a thin solid line. All properties are drawn as function of radius. All lines are drawn as histograms; individual gridpoints are visible in the innermost part of the envelope. The central-star mass is indicated in the panel. The large density trough at $10^{18}\,$cm is due to the last thermal pulse $40\,000$ years before the star left the AGB.\label{fig3}}
\end{figure}

\subsection{Detailed physical model properties}\label{sec:modelprops}
In the initial model, we assumed mean Galactic-disk PNe abundances \citep[\zgd, these were first quoted by][]{PeKiSc.:98} for nine elements (see Table~\ref{sandint3}); except for values of carbon and nitrogen, {\zgd} is close to solar. The abundance values $\epsilon_{i}$ are specified in both logarithmic number fractions relative to hydrogen, i.e., $\epsilon_{i}=\log\,(n_{i}/n_{\text{H}})+12$, and as mass fraction percentages $\beta_{i}=A_{i}X_{\text{H}}n_{i}/n_{\text{H}}$, where $A_i$ is the atomic mass relative to hydrogen and $X_{\text{H}}$ the hydrogen mass fraction. Other properties of the initial model are the central-star mass $M=0.595\,\msun$, the effective temperature $\teff=5865\,$K, the luminosity $L=5601\,\lsun$, and properties of the slow AGB wind such as the expansion velocity $v_{\infty}=10\,\kms$ \citep[see][]{StSzSc:98}. The density and the velocity structures of the initial AGB wind-envelope model are shown in Fig.~\ref{fig3}.

Immediately at the start of the RHD calculations, the abundances of the CS fast wind are replaced with a second set of abundances, {\zwr}. Assuming empirically determined abundances of the object {\BD}, we based these abundances on values of \citet[see table~2 therein]{MaHidAr.:07}. The CS fast wind adds mass at the inner boundary of the grid.

\begin{table}
\caption{Fast-wind model parameters of Wolf-Rayet stars.}
\label{sandint4}
\tabcolsep=4.5pt
\begin{tabular}{rrrrrrrl}
\hline\hline\\[-2ex]
\noalign{\smallskip}
\multicolumn{1}{r}{Id.}
     & \multicolumn{1}{c}{Object}
     & \multicolumn{1}{c}{$\teff$}
     & \multicolumn{1}{c}{$\log\dot{M}$}
     & \multicolumn{1}{c}{$\log\dot{M}_{\text{s}}$}
     & \multicolumn{1}{c}{$v_{\infty}$}
     & \multicolumn{1}{c}{$\log L$}
     & \multicolumn{1}{l}{R.}\\[0.0ex]\hline\\[-1.8ex]
 1 & \object{V\,348\,Sgr}       &  20 & $-6.53$ & $-7.03$ &  190 & 3.70 & 1\\
 2 & \object{IRAS\,21282}       &  28 & $-6.98$ & $-7.52$ &  180 & 3.75 & 2\\
 3 & \object{Hen\,2-113}        &  30 & $-5.54$ & $-6.65$ &  200 & 4.51 & 1\\
 4 & \object{K\,2-16}           &  30 & $-6.36$ & $-6.86$ &  300 & 3.70 & 1\\
 5 & \object{M\,4-18}           &  31 & $-6.01$ & $-6.49$ &  350 & 3.67 & 1\\
 6 & \object{CPD$-$56$\degr$8032} &  32 & $-5.57$ & $-6.34$ &  240 & 4.06 & 1\\
 7 & \object{PM\,1-188}         &  35 & $-5.70$ & $-6.20$ &  360 & 3.70 & 2\\
 8 & \object{SwSt\,1}           &  40 & $-6.90$ & $-7.08$ &  400 & 3.27 & 3\\
 9 & \object{\BD}               &  48 & $-6.30$ & $-6.30$ &  700 & 4.00 & 4\\
10 & \object{He\,2-99}          &  49 & $-5.59$ & $-6.11$ &  900 & 3.72 & 1\\
11 & \object{M\,2-43}           &  65 & $-6.08$ & $-6.24$ &  850 & 3.24 & 2\\
12 & \object{NGC\,40}           &  73 & $-6.25$ & $-6.25$ & 1000 & 3.58 & 4\\
13 & \object{He\,2-459}         &  77 & $-5.01$ & $-6.01$ & 1000 & 4.37 & 2\\
\hline
\end{tabular}
\tablefoot{Column 1, identification number used in Fig.~\ref{fig4}; Col.~2, name of the star; Col.~3, effective temperature [K]; Cols.~4 and 5, original and scaled mass-loss rates $[\text{M}_{\sun}\text{yr}^{-1}]$; Col.~6, expansion velocity $[\text{km}\,\text{s}^{-1}]$; Col.~7, luminosity $[L_{\sun}]$; Col.~8, reference.}
\tablebib{(1) \citet{LeHaJe:96}; (2) \citet{LeHa:98}; (3) Todt et al. (in prep.); (4) \citet{MaHidAr.:07}.}
\end{table}

\begin{figure}
\centering
\includegraphics{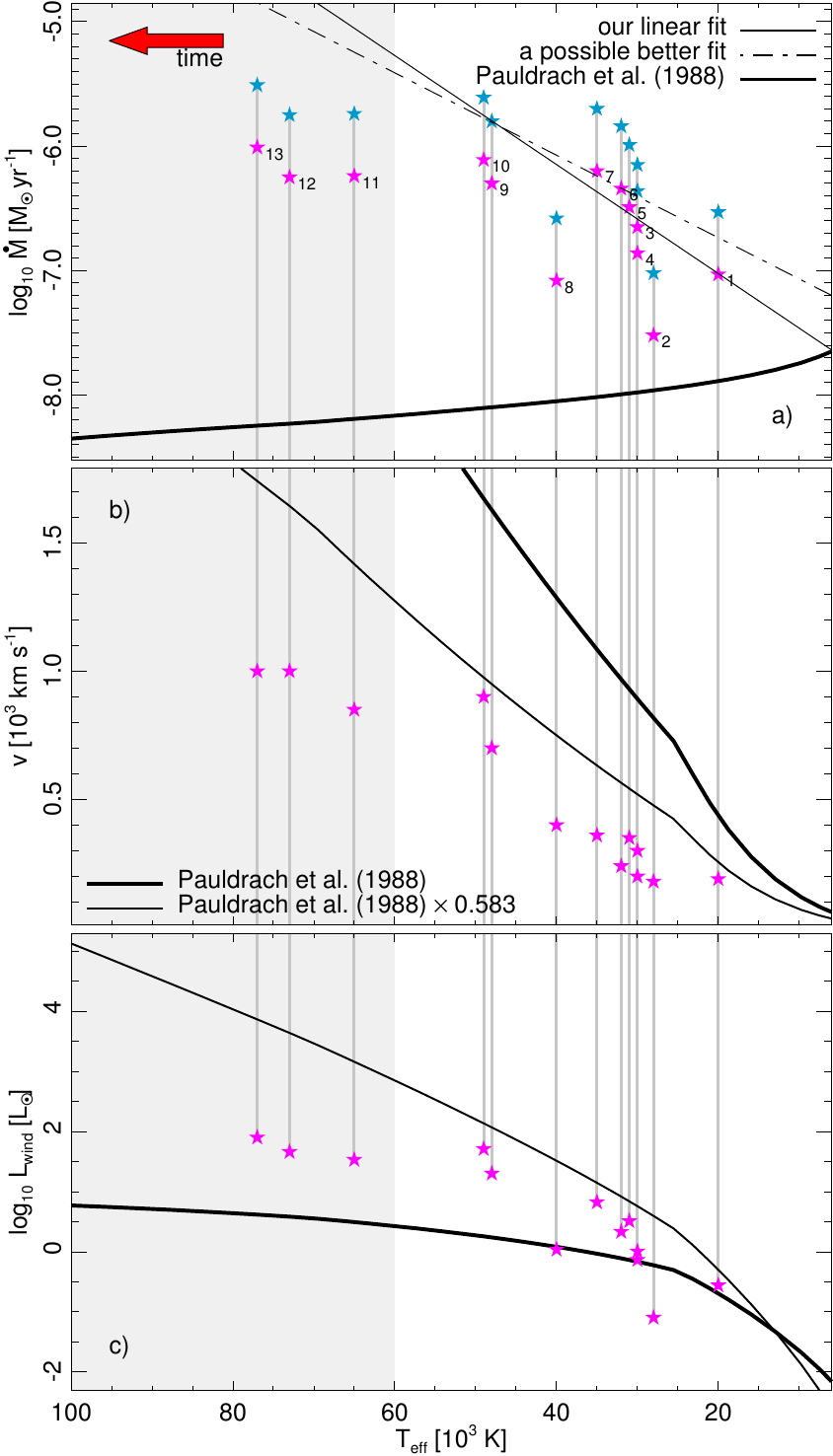}
\caption{Models, scaled models, and empirical values of fast winds of PNe: \textbf{a}) the mass-loss rate $\dot{M}$, \textbf{b}) the outflow velocity $v$, and \textbf{c}) the mechanical wind luminosity $L_{\text{wind}}$. Scaled values of $\dot{M}$ are illustrated with both magenta ($\mathcal{D}=10$) and blue ($\mathcal{D}=1$) star symbols. All symbols are indicated with both object (Table~\ref{sandint4}) and gray interconnecting lines for easy identification. Thick and thin lines indicate models based on \citet{PaPuKu.:88} and actually used relations, respectively. The shaded part where $\teff\ge60\,000\,$K indicates regions where both the models and the used relations fit empirical values poorly.\label{fig4}}
\end{figure}

\citet{PaPuKu.:88} present fast-wind models derived for hydrogen-rich conditions; the model mass-loss rate $\dot{M}$ decreases with time, while the outflow velocity $v$ increases with the effective temperature (i.e., with time). More recently, \citet{PaHoMe:04} find that the mass-loss rate instead appears to increase with time for $\teff\la50\,000\,$K. \citet{LeHaJe:96} present results of observations of CSPNe with hydrogen-deficient compositions, i.e., [WC] stars, where the mass-loss rate increases with time. Outflow velocities simultaneously increase less steeply than with a hydrogen-rich composition. Empirical values of effective temperatures, luminosities, and mass-loss rates are listed for thirteen Wolf-Rayet stars in Table~\ref{sandint4}. The table shows both mass-loss rates of the respective original reference and values that were rescaled, $\dot{M}_{\text{s}}$. The rescaled values use a standard luminosity $\log L=3.7\,\lsun$ and a clumping density ratio $\mathcal{D}=\rho_{\text{clumps}}/\rho=10$. The rescaling was made using the invariance of the so-called transformed radius $R_{\text{t}}$ \citep{ScHaWe:89,HaKo:98},
\begin{eqnarray}
R_{\text{t}}=R_*\left(\frac{v_{\infty}}{2500\,\kms}\middle/\frac{\dot{M}\mathcal{D}^{1/2}}{10^{-4}\,\msun\text{yr}^{-1}}\right)^{2/3},
\end{eqnarray}
where $R_*$ is a stellar photospheric radius ($R_*\sim L^{1/2}T_{\text{eff}}^{-2}$). This expression yields $\dot{M}\propto L^{3/4}\mathcal{D}^{-1/2}$ when $R_{\text{t}}$ and {\teff} are kept fixed. Mass-loss rates (assuming both $\mathcal{D}=10$ and $\mathcal{D}=1$), outflow velocities, and mechanical wind luminosities ($L_{\text{wind}}=\dot{M}v^2/2$) of both models and scaled models are shown in Fig.~\ref{fig4}, together with the model relations of Pauldrach et al.

The mass-loss descriptions of \citet{PaPuKu.:88} and \citet{PaHoMe:04} are both poor fits to the scaled empirical values of [WC] stars. Instead, we made a linear fit to the empirical mass-loss rates as function of effective temperature, where $\teff\la50\,000\,$K; we accounted for all objects but M\,2-43, NGC\,40, and He\,2-459. For the outflow velocities, we simply scaled the expression of \citet{PaPuKu.:88} with the factor 0.583; the scaled relation lies higher than the empirical values since we found the high velocities necessary to form a bubble at about $\teff\simeq47\,000\,$K, which is the currently measured temperature of \BD. Despite a flattening mass-loss rate as is seen in the measurements, the mechanical luminosity increases monotonically with time. The figure illustrates poor fits where $\teff\ga60\,000\,$K; we show this figure to emphasize the need for models that work with H-deficient abundances at low as well as high effective temperatures.

Heat conduction is always active during the RHD calculations. While our \textsc{nebel} code can use both numerical approaches HC1 and HC2 presented in {\rS} to account for heat flux below and above the saturation limit, we only used the HC2 limiter here (Eq.~\ref{sandineh2}). After the RHD calculations were started, the full model evolution was followed across the Hertzsprung-Russell diagram until the effective temperature reached $\teff\approx70\,000\,$K. We are only interested in the formation and early bubble evolution, which is why we did not consider the subsequent evolution.

\subsection{Comparing the HC and HC$_{\text{H}}$ models}\label{sec:models}
In comparison to the models that use an H-rich fast wind, models that use an H-deficient abundance composition in the fast wind experience two differences: the different abundances result in much more efficient cooling, and the mechanical luminosity of the fast wind is higher. More efficient cooling results in less available energy to form a bubble, and the PN evolution therefore changes \citep[also see][]{MeLu:02}. Here, we merely note that in our model the bubble starts forming when $\teff\simeq50\,000\,$K.

\begin{figure}[!ht]
\includegraphics{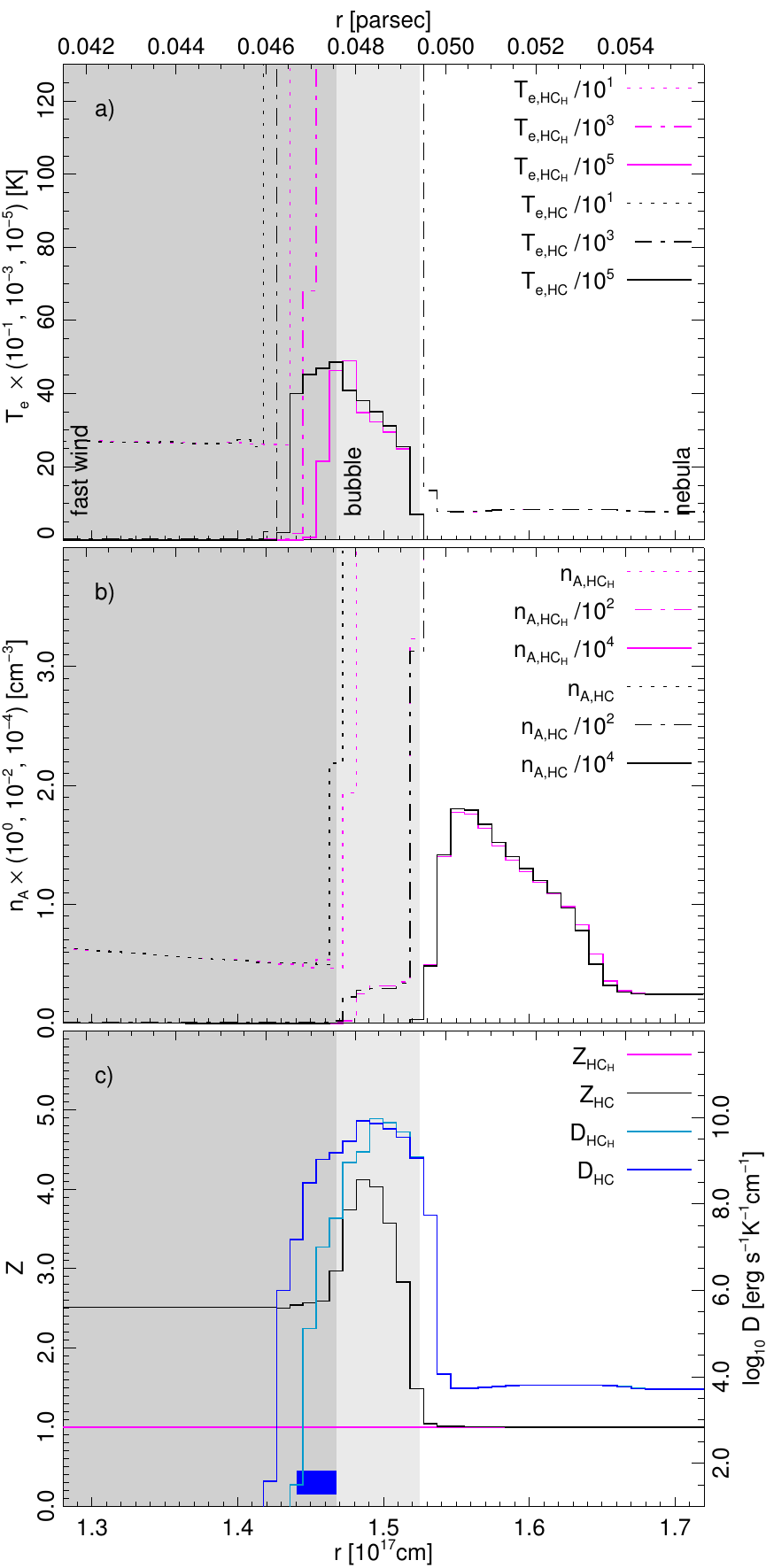}
\caption{Physical structure of PN when $\teff\simeq50\,500\,$K, the fast-wind speed $v=1031\,\kms$, and the model age is 4246\,yr: \textbf{a}) the electron temperature {\te}; \textbf{b}) the total ion density {\na}; \textbf{c}) the effective charge $Z$ and the diffusion coefficient $D$. The electron temperature and the total ion density are shown at three scales as indicated in each panel. Values of model HC (HC$_\text{H}$) are drawn with black (magenta) lines; for the diffusion coefficient the lines are blue (light blue). The heat-conduction flux of model HC is saturated in the region that is marked with a blue box. The chemical discontinuity between the H-rich nebula and the H-deficient inner parts is in the bubble-nebula interface. All lines are drawn as histograms to emphasize individual gridpoints. The fast wind (bubble) is drawn on a medium gray (light gray) background for model HC.\label{fig5}}
\end{figure}

\begin{figure}
\includegraphics{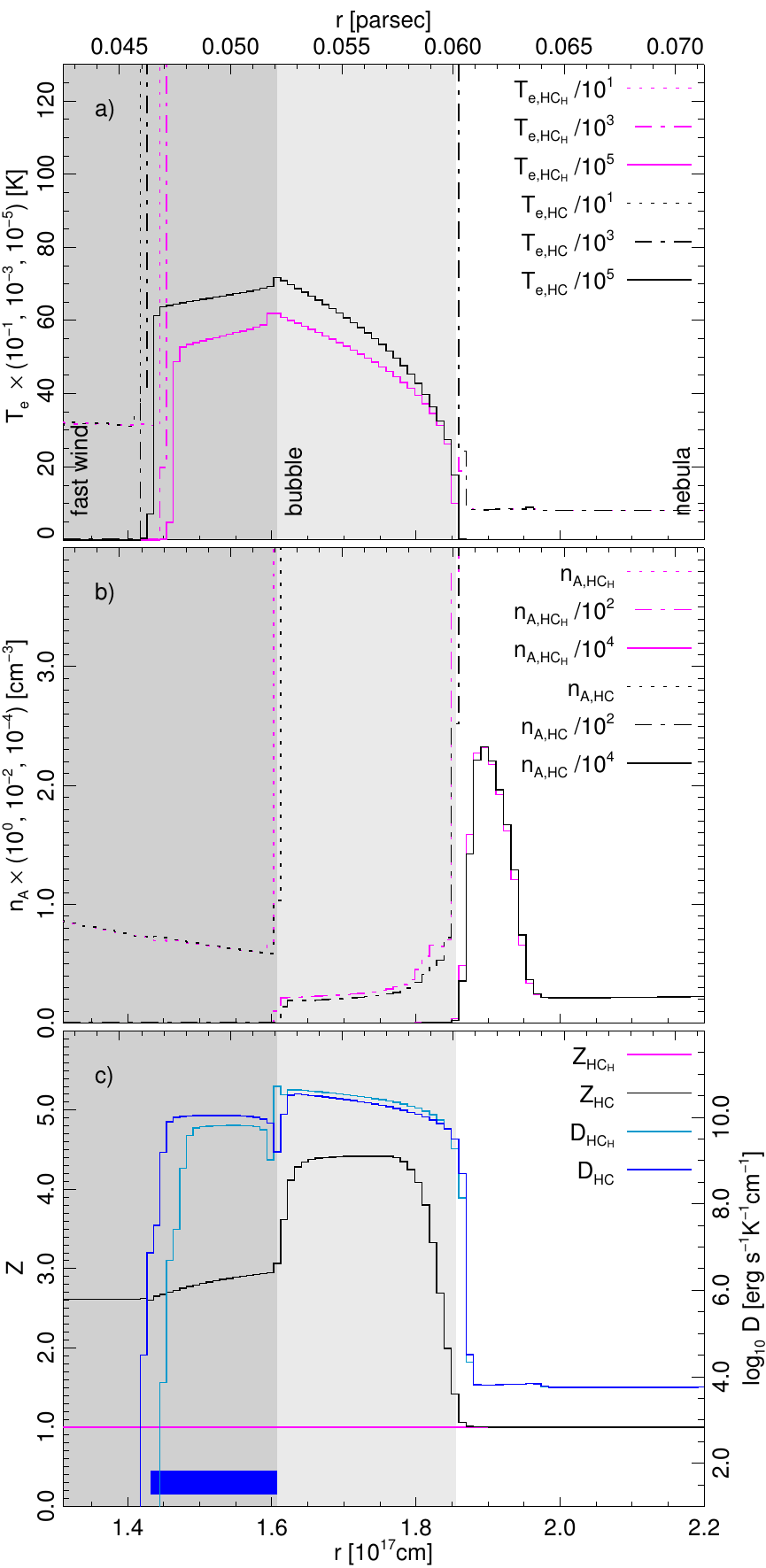}
\caption{Physical structure of PN when $\teff\simeq60\,000\,$K, the fast-wind speed $v=1154\,\kms$, and the model age is 4914\,yr. See Fig.~\ref{fig5} for more details.\label{fig6}}
\end{figure}

\begin{figure}
\includegraphics{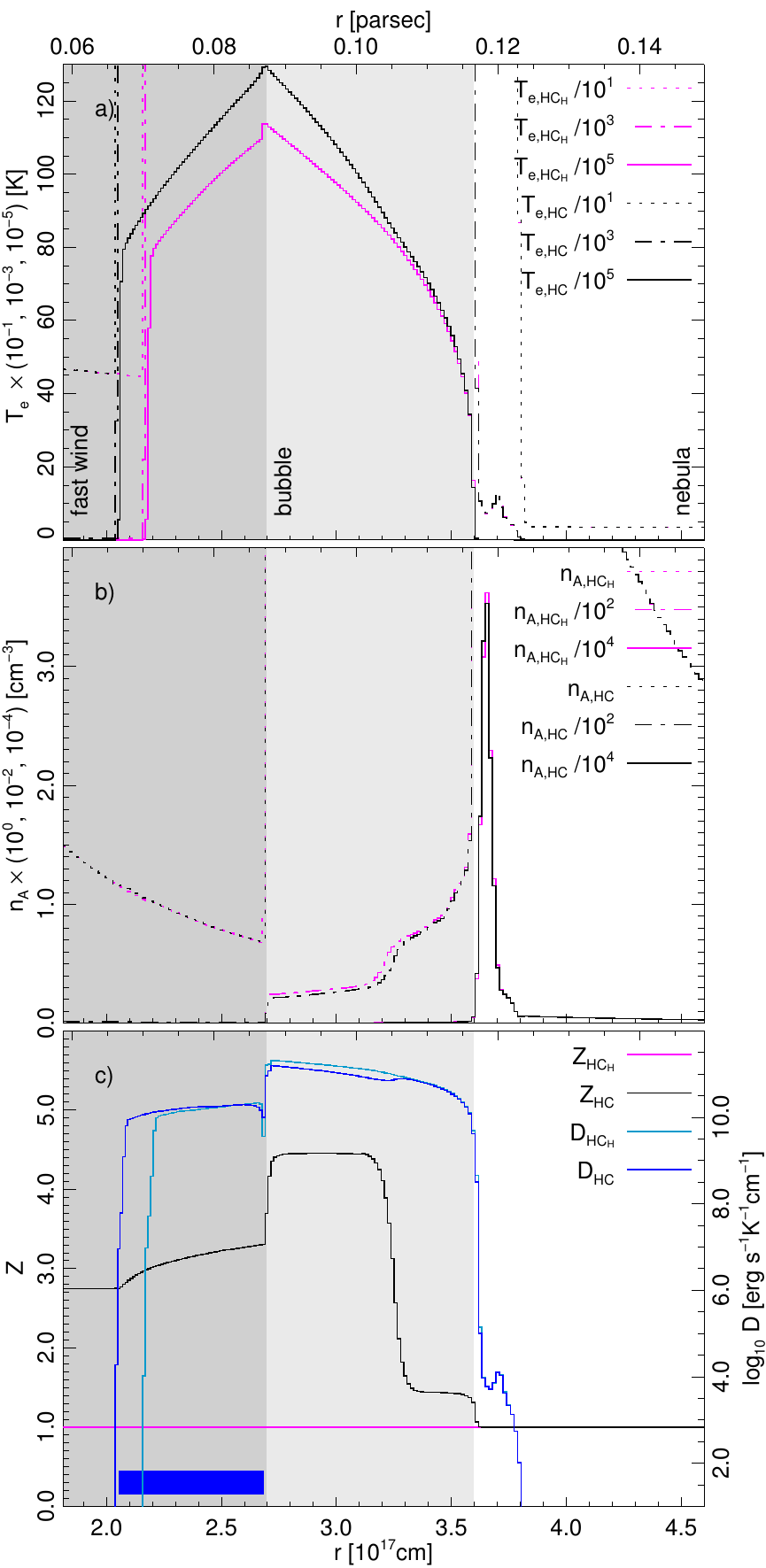}
\caption{Physical structure of PN when $\teff\simeq70\,000\,$K, the fast-wind speed $v=1575\,\kms$, and the model age is 5546\,yr. See Fig.~\ref{fig5} for more details.\label{fig7}}
\end{figure}

\begin{figure}
\includegraphics{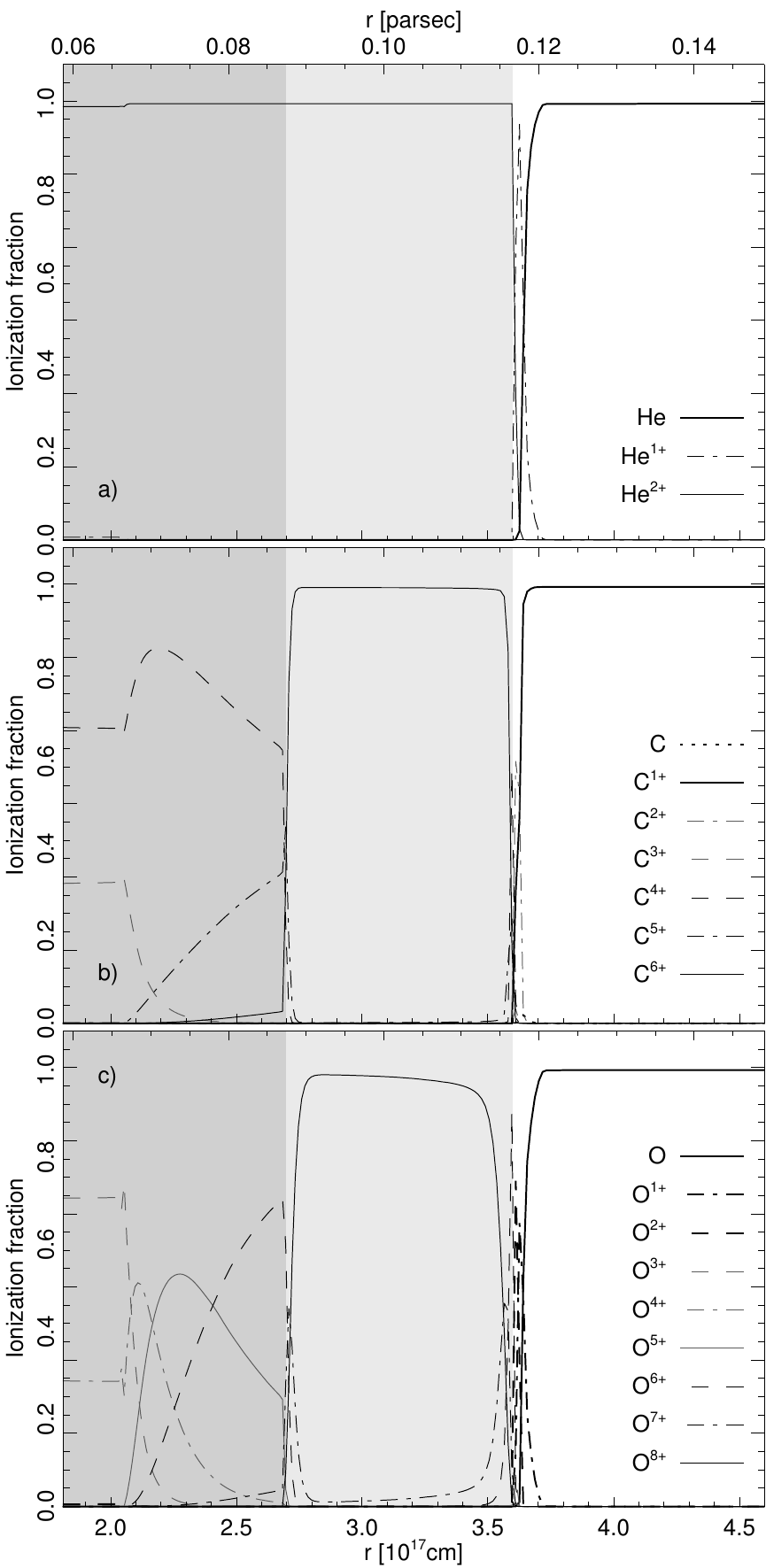}
\caption{For model HC at $\teff\simeq70\,000\,$K (Fig.~\ref{fig7}), the three panels show radial profiles of the ionization fractions of: \textbf{a)} helium, \textbf{b)} carbon, and \textbf{c)} oxygen. The fast wind (bubble) is drawn on a medium gray (light gray) background.\label{fig8}}
\end{figure}

\begin{figure}
\includegraphics{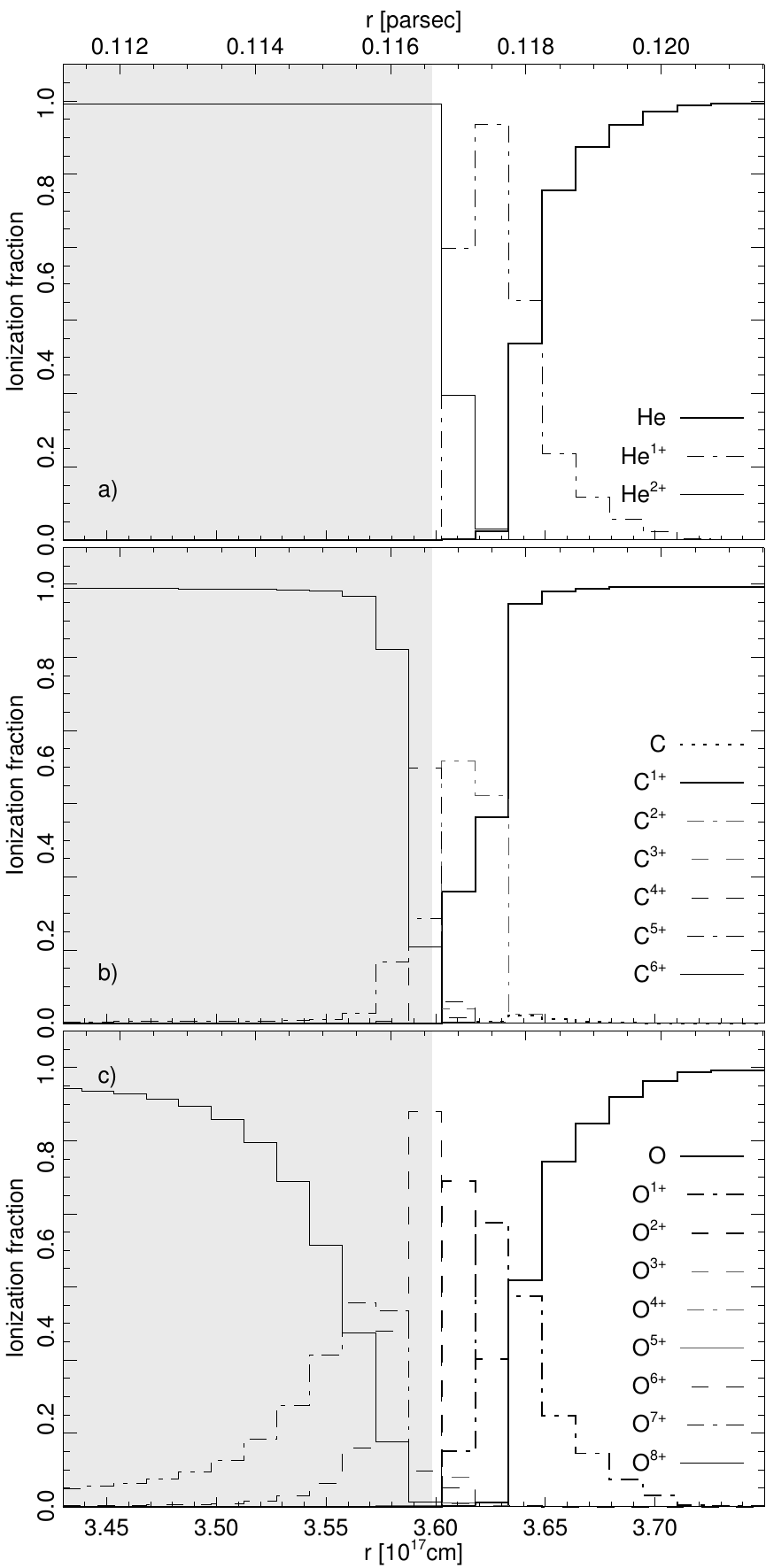}
\caption{Zoom in of Fig.~\ref{fig8} at the bubble-to-nebula interface. The histogram lines show gridpoints of individual ions as indicated in the respective panel.\label{fig9}}
\end{figure}

The physical structure of the inner regions of the PN is shown as radial ($r$) plots, which include the inner parts of the nebula, the bubble, and outer parts of the fast wind at three different stages during the bubble evolution: Fig.~\ref{fig5}) when the bubble forms at $\teff\simeq50\,500\,$K; Fig.~\ref{fig6}) at $\teff\simeq60\,000\,$K; and Fig.~\ref{fig7}) at $\teff\simeq70\,000\,$K (this case also shows all of the extent of the remaining nebula). Absciss{\ae} and ordinates use different scales. The ionization structures of the bubble at $\teff\simeq70\,000\,$K are shown for helium, carbon, and oxygen in Figs.~\ref{fig8} and \ref{fig9}.

The nebula already starts forming at $\teff\simeq13\,000\,$K. When the bubble starts forming -- here its range is $1.47\le r\le1.525\times10^{17}\,$cm -- at $\teff\simeq50\,000\,$K, the nebula already extends $1.525\le r\le2.6\times10^{17}\,$cm; the relative size of the bubble to the nebula is 5 per cent. The two models are very similar at this point. At this time, the outermost part of the fast wind has already reached the same temperature as the bubble. The density reaches a lower value of $\na=0.5\,\ccm$ in the fast wind. In the outer bubble, it increases to $\na\simeq300\,\ccm$. The density is the highest immediately ahead of the heat conduction front in the nebula, where $\na\la2\times10^4\,\ccm$. The temperature in the bubble (nebula) is 4.8--$2.5\times10^6\,$K (about 8000\,K). The effective charge $Z$ is markedly higher in the inner part of the bubble of the HC model than in the outer part. The diffusion coefficient is the property that differs the most between HC and HC$_{\text{H}}$, owing to its dependence on the effective charge and its high sensitivity to the temperature. However, heat conduction is negligible in the fast wind and the nebula when compared to the bubble (large temperature gradients are required to make it significant). Most of the bubble mass and energy are located at the front of the bubble where the two models are very similar.

Once the effective temperature has evolved to $\teff\simeq60\,000\,$K, the nebula extends $1.85\le r\le2.9\times10^{17}\,$cm, and the bubble $1.60\le r\le1.85\times10^{17}\,$cm. The models are still very similar except in the bubble. In the bubble, the HC model reaches $\te\simeq7\times10^6$K at the inner boundary, while the HC$_{\text{H}}$ model where $D_{\text{HC}_{\text{H}}}$ is higher than $D$ reaches $\te\simeq6\times10^6$K; the same differences are seen in the outermost part of the fast wind. The temperature differences are reflected in the diffusion coefficient. Because of the less efficient heat conduction of the HC model, the bubble is less efficiently cooled (which could also be due to lower density, and possibly to a slightly more efficient evaporation). The bubble density is slightly lower in the HC model (12--36 per cent). The hot region in the fast wind is slightly smaller in the HC$_{\text{H}}$ model -- exact details are unimportant as there is very little mass and energy in this region. The bubble has grown faster than the nebula; the bubble is now 30 per cent the size of the nebula.

At $\teff\simeq70\,000\,$K, the rim has nearly caught up with the shell, and the nebula is now a thin dense shell; the remaining nebula extends $3.6\le r\le3.8\times10^{17}\,$cm. The bubble extends over a much larger region, $2.7\le r\le3.6\times10^{17}\,$cm. The temperature difference is larger in this model, $T_{\text{e,HC}}\simeq13\times10^6\,$K versus $T_{\text{e,HC}_{\text{H}}}\simeq11\times10^6\,$K. However, the amount of mass in the inner part of the bubble is a small fraction of the total amount of mass of the bubble. The models are indistinguishable outside of the bubble; this indicates the small influence of the physical details on the PN evolution. The outermost part of the hot fast wind is larger for the HC model, as it cools this region less efficiently.

The lower heat conduction efficiency results in higher temperatures in the new heat conduction formulation. Higher temperatures lead to slightly steeper temperature gradients, which partly compensates for the lower efficiency. The heat conduction difference is smaller in the bubble front (Z closer to 1), which also explains the very similar outcome of HC and HC${_\text{H}}$. More nebular mass is evaporated and added to the bubble with more efficient heat conduction; the additional mass due to the evaporation increases the cooling capability of the bubble to further decrease its temperature; this explains the $10^6\,$K temperature difference between the two models. Finally, Figs.~\ref{fig6} and \ref{fig7} show how the outer part of the bubble after 632\,yr evolution has been filled with evaporated hydrogen-rich matter in the HC model (see Fig.~\ref{fig7} inside the contact discontinuity where $3.3\la r\la3.6\times10^{17}\,$cm).

The radial ionization structures of model HC for all ions of helium, carbon, and oxygen are shown in Fig.~\ref{fig8}; a zoom in of the interface between the bubble and the interface is shown in Fig.~\ref{fig9}. The figures show nearly fully ionized atoms in the bubble. More atoms assume lower ionization levels in the interface to the nebula, as expected from the lower temperature there.

\section{Influence of heat conduction on the diffuse X-ray emission}\label{sec:Xray}
In the following, we investigate the observable consequences of the small differences in the physical structure of the hot bubble between models HC and HC$_{\text{H}}$. We computed the X-ray emission based on the different hydrodynamical hot bubble structures at selected positions along the evolutionary sequences for this purpose in a post-processing step.

In addition to the two previously discussed models, we have considered a third model, named HC$_{\text{H}}^\ast$, where the diffusion coefficient is computed exactly as in the case of HC$_{\text{H}}$, but with a global scaling factor of $1/2$ -- $D($HC$_{\text{H}}^\ast) = D($HC$_{\text{H}})/2$. The setup of this simulation is otherwise identical to that of the other two models. The scaling factor of 1/2 roughly corresponds to the ratio $D/D_{\text{\rS}}$ at $Z=4$ (ionized carbon). The heat conduction efficiency of model HC is in this sense intermediate between models HC$_{\text{H}}$ and HC$_{\text{H}}^\ast$.

As described in more detail in {\rS}, we used the well-documented \textsc{chianti} code \citep[][in version 6.0 \citealt{DeLaYo.:09}]{DeLaMa.:97}, which has been used extensively in various astrophysical contexts. In a first step, the X-ray emission was calculated individually for all spherical shells of a given model, where $\te > 10^5$~K (the contribution of cooler regions is negligible in the X-ray spectral range). If the shell is hydrogen rich, we used the abundance mix {\zgd} (see Table~\ref{sandint3}) in the calculation of the synthetic X-ray spectrum, otherwise we assumed a chemical composition similar to {\zwr}\footnote{The mass fraction of H and N is $2\times 10^{-2}$ and $1\times 10^{-5}$, respectively.}. The slight inconsistency between the chemical composition used in the hydrodynamical simulations and the a posteriori calculation of the X-ray emission is considered irrelevant to our strictly differential comparison.

For given temperature and particle densities, the synthetic spectra include contributions of emission lines of all considered elements and various continua (free-free, free-bound, and two-photon continuum). Since the X-ray emission is optically thin, the total X-ray spectrum emitted by the hot bubble can be computed by adding up the volume-weighted emission of the individual shells. In the present context, we define the X-ray luminosity $L_{\text{X}}$ (erg\,s$^{-1}$) as the total X-ray emission of the hot bubble in the energy range 0.3--2.0~keV (6.2--41.3~\AA). For the same energy band, we also computed the characteristic X-ray temperature $T_{\text{X}}$, which is the emissivity-weighted temperature of the X-ray emitting region, as defined in {\rS} (Eq.\,17).

\begin{figure}
\includegraphics[width=\columnwidth]{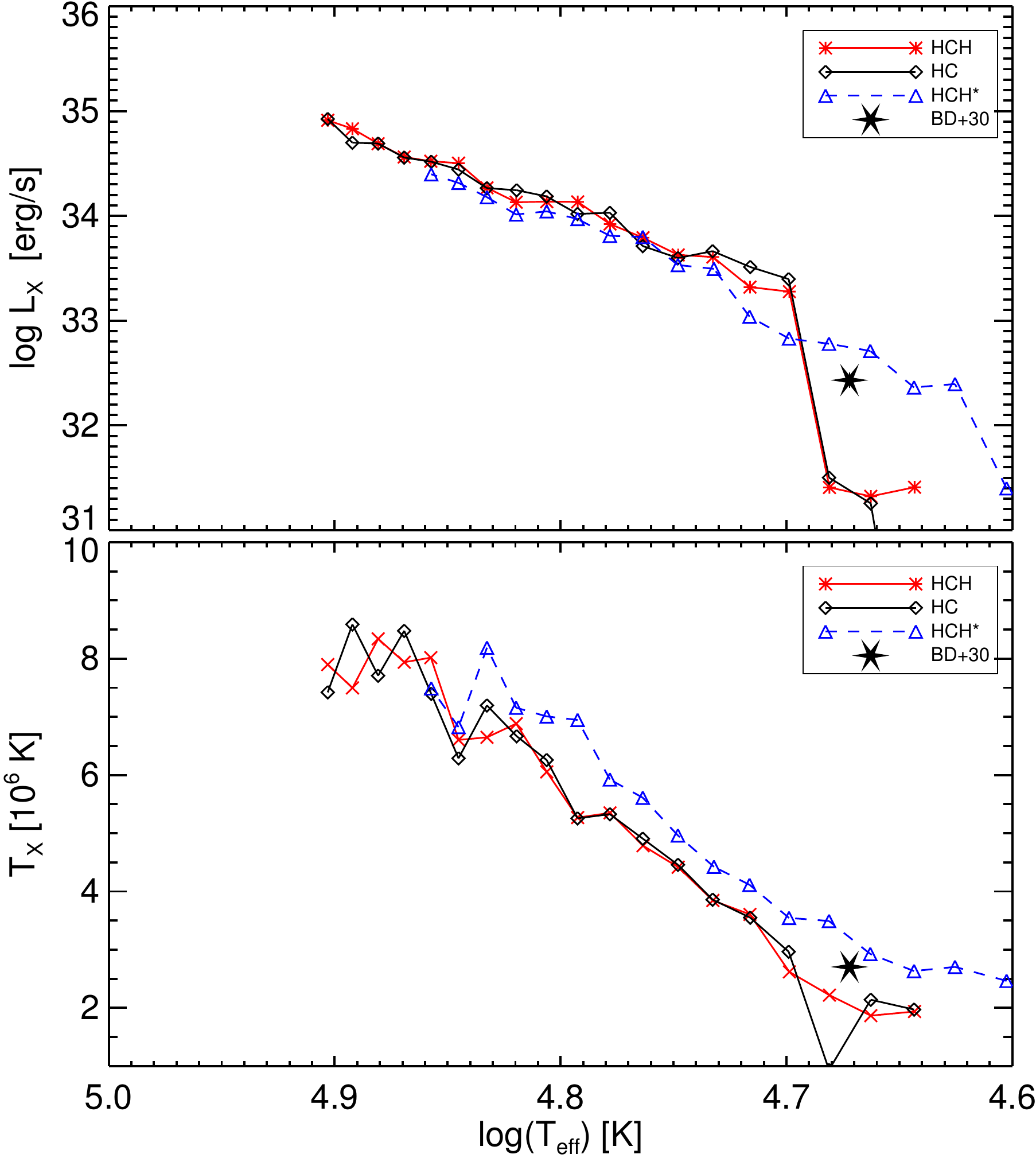}
\caption{Evolution of the synthetic X-ray luminosity $L_{\text{X}}$ (top panel) and temperature $T_{\text{X}}$ (bottom panel) integrated over the energy range 0.3--2.0\,keV versus {\teff} (which increases with time) for the three different models HC (black diamonds), HC$_{\text{H}}$ (red crosses), and HC$_{\text{H}}^\ast$ (blue triangles). The observed position of the PN {\BD} with a WR-type CS is indicated by a black star for comparison.}
\label{fig10}
\end{figure}

We show the evolution of the synthetic X-ray luminosity $L_{\text{X}}$ and temperature $T_{\text{X}}$ versus {\teff} (which increases with time) for the three different models in Fig.~\ref{fig10}. We conclude from these results that the X-ray properties of models HC$_{\text{H}}$ and HC are indistinguishable within the uncertainties of the numerical models. In addition, our experiment shows that while the provision of a $Z$-dependent diffusion coefficient hardly influences the X-ray properties of the hot bubble, a global scaling of the pure hydrogen diffusion coefficient by a factor 1/2 has a clearly noticeable impact. In the latter case, the bubble forms earlier in the PN evolution, namely at the time $t\approx 3580$\,yr when the central star has reached an effective temperature of $\teff\approx42\,000$\,K and a wind speed of $v\approx800\,\kms$. In the case of the HC$_{\text{H}}$ and HC models, the hot bubble forms significantly later, at  $t\approx 4250$\,yr, when $\teff\approx50\,500$\,K and $v\approx1000\,\kms$ (see Fig.~\ref{fig5}). The bubble formation is associated with a steep increase of the X-ray luminosity, as is apparent in the upper panel of Fig.\,\ref{fig10} near $\log\teff=4.7$ and $\log\teff=4.6$, respectively.

It is also seen that the reduced heat conduction of HC$_{\text{H}}^\ast$ results in a slightly lower X-ray luminosity and a significantly higher X-ray temperature at any given time. This behavior is in qualitative agreement with our findings in Paper~V that heat conduction leads to a delayed formation of the central hot bubble, while enhancing the X-ray luminosity and reducing the characteristic X-ray temperature of its diffuse X-ray emission. However, the increase of $L_{\text{X}}$ and $T_{\text{X}}$ with {\teff} is much steeper here than found previously for similar CS masses \citep[see Fig.~18 in {\rS}, and Figs.~5 and 6 in][]{RuChGr.:13}; this is attributed to the different wind model used in the present work, where the wind luminosity attains much higher values over the same time period (see Fig.~\ref{fig4}).

\begin{figure}
\includegraphics[width=\columnwidth]{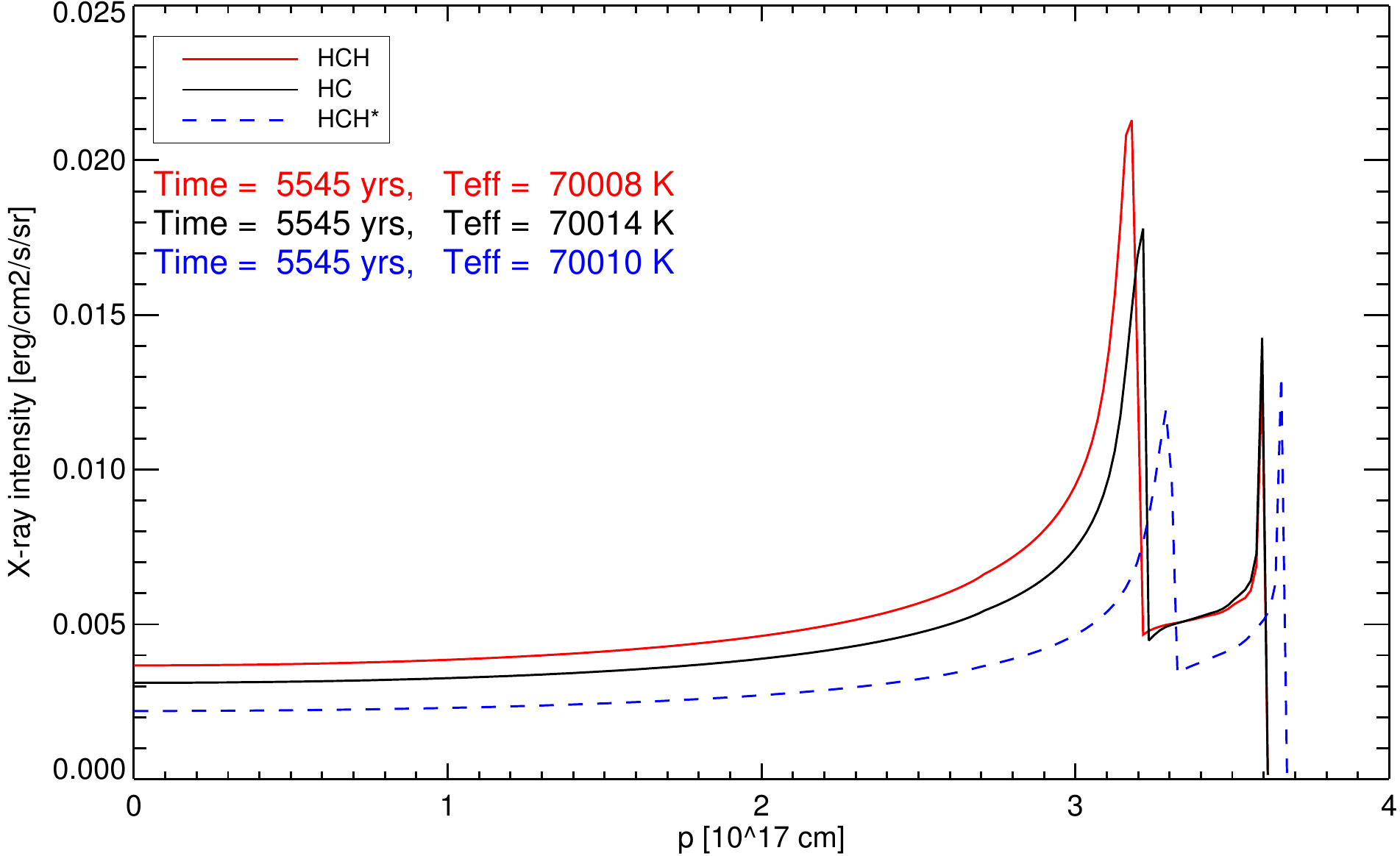}
\caption{Synthetic (unabsorbed) X-ray surface brightness integrated over the energy range 0.3--2.0\,keV versus the impact parameter $p$ for the three models HC (black solid line), HC$_{\text{H}}$ (red solid line), and HC$_{\text{H}}^\ast$ (blue dotted line) at $t=5545$\,yr and $\teff\approx70\,000$\,K.}
\label{fig11}
\end{figure}

We have also used our models to compute the X-ray surface-brightness distribution in the plane of the sky (in $\text{erg}\,\text{cm}^{-2}\text{s}^{-1}\text{sr}^{-1}$) in the same energy band as above. The X-ray emissivity is for this purpose integrated along lines of sight perpendicular to the plane of the sky for a series of impact parameters, $p$. Figure\,\ref{fig11} shows the radial profile of the X-ray surface brightness for the three heat-conduction models at the end of the considered time evolution ($t=5545$\,yr) when the CS has reached $\teff\approx70\,000$\,K.

The chemical discontinuity is clearly visible as a jump in the radial intensity profile near $r=3.2\times 10^{17}$\,cm. The hydrogen-deficient inner bubble appears slightly more extended in the case of HC than in HC$_{\text{H}}$. Presumably, this is a consequence of the somewhat lower evaporation rate in HC. The higher bubble temperature (see Fig.\,7) may also play a role. Together with the slightly lower density, this leads to a somewhat reduced overall X-ray luminosity of HC. We cannot see a difference in the radial expansion rate of the outer edge of the hot bubble between the HC and HC$_{\text{H}}$, as all physical parameters (including the diffusion coefficient) are very similar in the hydrogen-rich outer part of the hot bubble (Fig.~\ref{fig7}). The inner wind shock at $r=2.7\times 10^{17}$\,cm is not seen in X-rays owing to the markedly low density in the inner bubble.

The total radial thickness of the HC$_{\text{H}}^\ast$ model is distinctly larger than in the other two models because the bubble of the HC$_{\text{H}}^\ast$ model formed earlier and has had more time to grow. As a result of the general reduction of the diffusion coefficient by a factor 2, the overall temperature of the hot bubble is higher, but the total particle density is lower, such that the gas pressure inside the bubble is the same in all three cases. Remarkably, the thickness of the hydrogen-rich outer part of the bubble, which is the result of the evaporation of cool nebular material, is not significantly reduced in the case of HC$_{\text{H}}^\ast$ despite the lower heat conduction efficiency. The higher temperatures plausibly lead to a slightly steeper temperature gradient at the evaporation front at the outer edge of the hot bubble, which partly compensates for the lower diffusion coefficient assumed in the HC$_{\text{H}}^\ast$ scheme.

A comparison of our model predictions with the observed X-ray luminosity of {\BD} -- which is a PN with a WR-type central star -- indicates that our current model is an inadequate representation of this object. As is demonstrated in Fig.~\ref{fig10}, the formation of the hot bubble obviously takes place too late (at too high {\teff} and $v$) in the HC and HC$_{\text{H}}$ models. In contrast, the model with reduced heat conduction, HC$_{\text{H}}^\ast$, appears to work much better. However, it is premature to conclude that the heat conduction schemes HC and HC$_{\text{H}}$ are unrealistic since other factors also influence the circumstances of the hot bubble formation. These factors include the kinematics and the detailed chemical composition (cooling function) of the fast wind. The detailed modeling of {\BD} is beyond the scope of this paper.

\section{Conclusions}\label{sec:conclusions}
We extended our radiation hydrodynamic PN models with a formulation of heat conduction that allows the use of any chemistry. Through this physics extension, we can now model chemically stratified plasmas such as PNe that show hydrogen-deficient hot bubbles.

The heat conduction equation is described by a diffusion coefficient multiplied by a temperature gradient where the chemistry is represented by additional terms in the diffusion coefficient that depend on the effective charge in the plasma. The higher effective charge in a hydrogen-deficient plasma results in a somewhat lower diffusion coefficient than when there is only hydrogen. In the case of PNe, a lower diffusion coefficient simultaneously results in a somewhat higher temperature and a slightly steeper temperature gradient near the interface between the hot bubble and the surrounding cooler nebula. Differences in the heat conduction efficiency between models using the new formulation and the old formulation assuming a pure hydrogen composition are therefore very small. However, the largest deviation is found in the hot bubble, where there is very little mass. We compared two different theories of the heat conduction terms: the Fokker-Planck equation (\rSp) and the Chapman-Enskog-Burnett theory \citep{De:66,De:67a,De:67b}. We have found that the differences in outcome between the two theories are marginal, and so we used the simpler approach of {\rSp}.

We modeled a chemically stratified PN using both the new chemistry-dependent formulation and the old hydrogen-rich formulation of heat conduction. The model consists of a regular hydrogen-rich slow AGB wind where a hydrogen-deficient fast wind starts blowing out into the slow wind when the computations begin. The result is a PN with a nebula outside of a hot bubble. We used empirical measurements of [WC]-type PNe to describe the mass-loss and velocity evolutions of the fast wind. The model parameters were otherwise chosen to be similar to those of a real object where a hot bubble forms when the effective temperature of the central star is $\teff=50\,000$K. We found that the two models evolve nearly identically. The only differences are seen in the hot and tenuous bubble, where the model that uses the chemistry-dependent formulation of heat conduction is up to $2\times10^6\,$K hotter than before.

We also calculated observational properties for the models in form of the synthetic X-ray temperature, luminosity, and surface-brightness distribution in the plane of the sky. Here, we also computed a third model that uses the old pure-hydrogen formulation, but where the diffusion-coefficient is half as high throughout the model domain, to mimic the effect the new-formulation diffusion coefficient inside the hydrogen-poor bubble has on the entire model. The X-ray properties of the first two models are seen to be indistinguishable, while the third model results in a somewhat higher temperature, lower luminosity, and a larger bubble.

In comparison to {\BD}, the hot bubble in our models forms too late at an excessive effective temperature and velocity of the fast wind. These models are therefore inadequate representations of this PN. While we improved the description of the heat conduction, there are several other factors that determine the model evolution. The physical properties of the fast wind, including the mass-loss and velocity evolutions, and the cooling function are such factors, whose descriptions need to be improved further to help the modeling of this and other PNe.

\begin{acknowledgements}
We thank W.-R. Hamann, L. Oskinova, and H. Todt (Univ. Potsdam) for fruitful and lively discussions on heat conduction in the context of hydrogen-deficient PNe. C.S. was supported by funds of DFG (SCHO-394/29-1), Land Brandenburg (SAW funds from WGL), PTDESY-05A12BA1, and the BMBF VIP program 03V0843. CHIANTI is a collaborative project involving George Mason University, the University of Michigan (USA), and the University of Cambridge (UK). 
\end{acknowledgements}


\begin{thebibliography}{34}
\expandafter\ifx\csname natexlab\endcsname\relax\def\natexlab#1{#1}\fi

\bibitem[{{Balick} \& {Frank}(2002)}]{BaFr:02}
{Balick}, B. \& {Frank}, A. 2002, \araa, 40, 439

\bibitem[{{Borkowski} {et~al.}(1990){Borkowski}, {Balbus}, \&
  {Fristrom}}]{BoBaFr:90}
{Borkowski}, K.~J., {Balbus}, S.~A., \& {Fristrom}, C.~C. 1990, \apj, 355, 501

\bibitem[{{Cohen} {et~al.}(1950){Cohen}, {Spitzer}, \& {Routly}}]{CoSpRo:50}
{Cohen}, R.~S., {Spitzer}, L., \& {Routly}, P.~M. 1950, Physical Review, 80,
  230

\bibitem[{{Cowie} \& {McKee}(1977)}]{CoMc:77}
{Cowie}, L.~L. \& {McKee}, C.~F. 1977, \apj, 211, 135

\bibitem[{{Dere} {et~al.}(1997){Dere}, {Landi}, {Mason}, {Monsignori Fossi}, \&
  {Young}}]{DeLaMa.:97}
{Dere}, K.~P., {Landi}, E., {Mason}, H.~E., {Monsignori Fossi}, B.~C., \&
  {Young}, P.~R. 1997, \aaps, 125, 149

\bibitem[{{Dere} {et~al.}(2009){Dere}, {Landi}, {Young}, {Del Zanna},
  {Landini}, \& {Mason}}]{DeLaYo.:09}
{Dere}, K.~P., {Landi}, E., {Young}, P.~R., {et~al.} 2009, \aap, 498, 915

\bibitem[{{Devoto}(1966)}]{De:66}
{Devoto}, R.~S. 1966, Physics of Fluids, 9, 1230

\bibitem[{{Devoto}(1967{\natexlab{a}})}]{De:67b}
{Devoto}, R.~S. 1967{\natexlab{a}}, Physics of Fluids, 10, 2105

\bibitem[{{Devoto}(1967{\natexlab{b}})}]{De:67a}
{Devoto}, R.~S. 1967{\natexlab{b}}, Physics of Fluids, 10, 354

\bibitem[{{Freeman} {et~al.}(2014){Freeman}, {Montez}, {Kastner}, {Balick},
  {Frew}, {Jones}, {Miszalski}, {Sahai}, {Blackman}, {Chu}, {De Marco},
  {Frank}, {Guerrero}, {Lopez}, {Zijlstra}, {Bujarrabal}, {Corradi},
  {Nordhaus}, {Parker}, {Sandin}, {Sch{\"o}nberner}, {Soker}, {Sokoloski},
  {Steffen}, {Toal{\'a}}, {Ueta}, \& {Villaver}}]{FrMoKa.:14}
{Freeman}, M., {Montez}, Jr., R., {Kastner}, J.~H., {et~al.} 2014, \apj, 794,
  99

\bibitem[{{Hamann} \& {Koesterke}(1998)}]{HaKo:98}
{Hamann}, W.-R. \& {Koesterke}, L. 1998, \aap, 335, 1003

\bibitem[{{Kastner} {et~al.}(2012){Kastner}, {Montez}, {Balick}, {Frew},
  {Miszalski}, {Sahai}, {Blackman}, {Chu}, {De Marco}, {Frank}, {Guerrero},
  {Lopez}, {Rapson}, {Zijlstra}, {Behar}, {Bujarrabal}, {Corradi}, {Nordhaus},
  {Parker}, {Sandin}, {Sch{\"o}nberner}, {Soker}, {Sokoloski}, {Steffen},
  {Ueta}, \& {Villaver}}]{KaMoBa.:12}
{Kastner}, J.~H., {Montez}, Jr., R., {Balick}, B., {et~al.} 2012, \aj, 144, 58

\bibitem[{{Leuenhagen} {et~al.}(1996){Leuenhagen}, {Hamann}, \&
  {Jeffery}}]{LeHaJe:96}
{Leuenhagen}, U., {Hamann}, W., \& {Jeffery}, C.~S. 1996, \aap, 312, 167

\bibitem[{{Leuenhagen} \& {Hamann}(1998)}]{LeHa:98}
{Leuenhagen}, U. \& {Hamann}, W.-R. 1998, \aap, 330, 265

\bibitem[{{Marcolino} {et~al.}(2007){Marcolino}, {Hillier}, {de Araujo}, \&
  {Pereira}}]{MaHidAr.:07}
{Marcolino}, W.~L.~F., {Hillier}, D.~J., {de Araujo}, F.~X., \& {Pereira},
  C.~B. 2007, \apj, 654, 1068

\bibitem[{{Mellema} \& {Lundqvist}(2002)}]{MeLu:02}
{Mellema}, G. \& {Lundqvist}, P. 2002, \aap, 394, 901

\bibitem[{{Montez} {et~al.}(2015){Montez}, {Kastner}, {Balick}, {Behar},
  {Blackman}, {Bujarrabal}, {Chu}, {Corradi}, {De Marco}, {Frank}, {Freeman},
  {Frew}, {Guerrero}, {Jones}, {Lopez}, {Miszalski}, {Nordhaus}, {Parker},
  {Sahai}, {Sandin}, {Schonberner}, {Soker}, {Sokoloski}, {Steffen},
  {Toal{\'a}}, {Ueta}, {Villaver}, \& {Zijlstra}}]{MoKaBa.:15}
{Montez}, Jr., R., {Kastner}, J.~H., {Balick}, B., {et~al.} 2015, \apj, 800, 8

\bibitem[{{Pauldrach} {et~al.}(1988){Pauldrach}, {Puls}, {Kudritzki}, {Mendez},
  \& {Heap}}]{PaPuKu.:88}
{Pauldrach}, A., {Puls}, J., {Kudritzki}, R.~P., {Mendez}, R.~H., \& {Heap},
  S.~R. 1988, \aap, 207, 123

\bibitem[{{Pauldrach} {et~al.}(2004){Pauldrach}, {Hoffmann}, \&
  {M\'endez}}]{PaHoMe:04}
{Pauldrach}, A.~W.~A., {Hoffmann}, T.~L., \& {M\'endez}, R.~H. 2004, \aap, 419,
  1111

\bibitem[{{Perinotto} {et~al.}(1998){Perinotto}, {Kifonidis}, {Sch\"onberner},
  \& {Marten}}]{PeKiSc.:98}
{Perinotto}, M., {Kifonidis}, K., {Sch\"onberner}, D., \& {Marten}, H. 1998,
  \aap, 332, 1044

\bibitem[{{Perinotto} {et~al.}(2004){Perinotto}, {Sch{\"o}nberner}, {Steffen},
  \& {Calonaci}}]{PeScSt.:04}
{Perinotto}, M., {Sch{\"o}nberner}, D., {Steffen}, M., \& {Calonaci}, C. 2004,
  \aap, 414, 993

\bibitem[{{Ruiz} {et~al.}(2013){Ruiz}, {Chu}, {Gruendl}, {Guerrero}, {Jacob},
  {Sch{\"o}nberner}, \& {Steffen}}]{RuChGr.:13}
{Ruiz}, N., {Chu}, Y.-H., {Gruendl}, R.~A., {et~al.} 2013, \apj, 767, 35

\bibitem[{{Schmutz} {et~al.}(1989){Schmutz}, {Hamann}, \&
  {Wessolowski}}]{ScHaWe:89}
{Schmutz}, W., {Hamann}, W.-R., \& {Wessolowski}, U. 1989, \aap, 210, 236

\bibitem[{{Soker}(1994)}]{So:94}
{Soker}, N. 1994, \aj, 107, 276

\bibitem[{{Sp\"ath} \& {Meier}(1990)}]{SpMe:90}
{Sp\"ath}, H. \& {Meier}, J. 1990, {Eindimensionale Spline -- Interpolations
  Algorithmen} (Oldenbourg Verlag)

\bibitem[{{Spitzer}(1962)}]{Sp:62}
{Spitzer}, L. 1962, {Physics of Fully Ionized Gases}, 2nd edn. (John Wiley \&
  Sons, Inc.) (\rSp)

\bibitem[{{Spitzer} \& {H{\"a}rm}(1953)}]{SpHa:53}
{Spitzer}, L. \& {H{\"a}rm}, R. 1953, Physical Review, 89, 977

\bibitem[{{Steffen} {et~al.}(2014){Steffen}, {Hubrig}, {Todt}, {Sch{\"o}ller},
  {Hamann}, {Sandin}, \& {Sch{\"o}nberner}}]{StHuTo.:14}
{Steffen}, M., {Hubrig}, S., {Todt}, H., {et~al.} 2014, \aap, 570, A88

\bibitem[{{Steffen} {et~al.}(2008){Steffen}, {Sch{\"o}nberner}, \&
  {Warmuth}}]{StScWa:08}
{Steffen}, M., {Sch{\"o}nberner}, D., \& {Warmuth}, A. 2008, \aap, 489, 173 (\rS)

\bibitem[{{Steffen} {et~al.}(1998){Steffen}, {Szczerba}, \&
  {Sch{\"o}nberner}}]{StSzSc:98}
{Steffen}, M., {Szczerba}, R., \& {Sch{\"o}nberner}, D. 1998, \aap, 337, 149

\bibitem[{{Stute} \& {Sahai}(2006)}]{StSa:06}
{Stute}, M. \& {Sahai}, R. 2006, \apj, 651, 882

\bibitem[{{Toal{\'a}} \& {Arthur}(2014)}]{ToAr:14}
{Toal{\'a}}, J.~A. \& {Arthur}, S.~J. 2014, \mnras, 443, 3486

\bibitem[{{Ulmschneider}(1970)}]{Ul:70}
{Ulmschneider}, P. 1970, \aap, 4, 144

\bibitem[{{Weaver} {et~al.}(1977){Weaver}, {McCray}, {Castor}, {Shapiro}, \&
  {Moore}}]{WeMcCa.:77}
{Weaver}, R., {McCray}, R., {Castor}, J., {Shapiro}, P., \& {Moore}, R. 1977,
  \apj, 218, 377

\end{thebibliography}
\end{document}